\documentstyle[epsf]{mn}
\newif\ifAMStwofonts
\AMStwofontstrue

\title[Non-Maxwellian excitation of the helium resonance
lines]{Enhancement of the helium resonance lines in the solar
atmosphere by suprathermal electron excitation II: non-Maxwellian
electron distributions}
\author[G.R.~Smith]{G.R.~Smith\footnotemark \\
Department of Physics (Theoretical Physics), Oxford University, 1 Keble Road,
Oxford OX1 3NP, UK}
\date{}
\pagerange{}
\pubyear{}

\begin{document}

\maketitle

\label{firstpage}

\begin{abstract}
In solar {\sc euv} spectra the He~{\sc i} and He~{\sc ii} resonance
lines show unusual behaviour and have anomalously high intensities
compared with other transition region lines. The formation of the
helium resonance lines is investigated through extensive non-LTE
radiative transfer calculations. The model atmospheres of Vernazza,
Avrett \& Loeser (1981) are found to provide reasonable matches to the
helium resonance line intensities but significantly over-estimate the
intensities of other transition region lines. New model atmospheres
have been developed from emission measure distributions derived by
Macpherson \& Jordan (1999), which are consistent with {\em SOHO}
observations of transition region lines other than those of
helium. These models fail to reproduce the observed helium
resonance line intensities by significant factors. 
The possibility that non-Maxwellian electron
distributions in the transition region might lead to 
increased collisional excitation rates in the helium lines is studied. 
Collisional excitation and ionization rates are re-computed for
distribution functions with power law suprathermal tails 
which may form by the transport of fast electrons from
high temperature regions. 
Enhancements of the helium resonance line intensities are
found, but many of the predictions of the models regarding line ratios
are inconsistent with observations. These results suggest that any
such departures from Maxwellian electron distributions are not
responsible for the helium resonance line intensities. 
\end{abstract}


\section{Introduction}
\footnotetext{E-mail: g.smith2@physics.ox.ac.uk}
The resonance lines of He~{\sc i} and He~{\sc ii} show unusual
behaviour when compared with other strong emission
lines in solar {\sc euv} spectra. Recent results from the
Solar and Heliospheric Observatory ({\em SOHO}) show that their
intensities in coronal holes are factors of 1.5--2.0 smaller than in
the quiet Sun (Peter 1999; Jordan, Macpherson \& Smith 2001), while
other lines formed at similar 
temperatures show only very small reductions in intensity. 
In the quiet Sun, the helium resonance line intensities are at least
an order of magnitude too large to be reproduced by emission measure
distributions that account for other transition region (TR) lines
(Macpherson \& Jordan 1999 -- hereafter MJ99). For discussions of
earlier observations of the helium resonance lines and attempts to
explain them see Hammer \shortcite{rh97} and MJ99. 
The discrepancies between observations and modelling of the helium
lines in the quiet Sun have been taken to imply that some process
preferentially enhances the helium resonance line intensities with
respect to other lines formed in the transition region in the quiet 
Sun, and that the enhancement is reduced in coronal holes. 
 
Current evidence suggests that in the quiet Sun the He~{\sc ii}, and
to a lesser extent the He~{\sc i}, resonance lines are formed mainly by
collisional excitation (see Smith \& Jordan 2002 -- hereafter Paper I -- for
supporting references). Because the helium
resonance lines have unusually large values of $W/kT_{\textrm{e}}$,
where $W$ is the excitation energy and $T_{\textrm{e}}$ is the electron
temperature, their collisional contribution
functions are sensitive to excitation by suprathermal
electrons. He~{\sc i} and He~{\sc ii} also have long ionization times
compared with other transition region species, which may allow
departures from ionization equilibrium in helium. Any process exposing
helium ions to larger populations of suprathermal electrons than in
equilibrium will tend to increase the collisional excitation rates of
the helium lines, while lines with smaller $W/kT_{\textrm{e}}$ will be
relatively unaffected. MJ99 reported that attempts to model such
effects had been generally unsuccessful in explaining the helium
resonance line emission \emph{relative} to other transition region
(TR) lines. While a conclusive explanation is still lacking, some
recent work appears promising (see Paper I). 

Fontenla, Avrett \& Loeser \shortcite{fal02} have extended their
earlier work \cite{fal93}, performing radiative transfer calculations
for hydrogen and helium including the effects of mass-conserving flows
as well as ambipolar diffusion and departures from ionization
equilibrium. His preliminary results show improved
agreement with observations, outflows (of 5--10 km s$^{-1}$
at $T = 10^{5}$ K) producing increases in intensity in the He~{\sc i}
and He~{\sc ii} resonance
lines of up to an order of magnitude. Inflow models produce
self-reversals in the He~{\sc i} 584.3-\AA\ line, whereas observations show
the line is more likely to be self-reversed in cell interior
regions (e.g.\ MJ99) where the line shows a \emph{blue} shift
\cite{hp99}. This implies that the models do not
yet include all relevant processes. As the models do not include the
effects of diffusion on any elements heavier than helium, it is also
uncertain as to how successful they will be in explaining the helium line
intensities \emph{relative} to other transition region lines.

Jordan (1975,1980) suggested two possible enhancement
mechanisms dependent on the magnitude of the temperature gradient that
might explain the coronal hole/quiet Sun contrast, Munro \&
Withbroe \shortcite{mw72} having found d$T/$d$h$ to be an order of
magnitude smaller inside coronal holes. The first involves the
transport of helium ions up the steep TR temperature gradient,
allowing them to be excited at electron temperatures higher than those
in equilibrium. Jordan (1980) found that intensity enhancements of up
to a factor of 5 could be produced in the He~{\sc ii} 303.8-\AA\ line
in the quiet Sun, and a more detailed study by Andretta et al.\ (2000)
suggests similar factors. A further investigation of the process using
an improved treatment of the excitation times of the helium ions is
reported in Paper I. The analysis was extended to include estimates of
the effects on the He~{\sc i} 584.3-\AA\ and 537.0-\AA\ lines. We
conclude, as did Andretta et al.\ \shortcite{aea00} for the He~{\sc
ii} line, that current observations are consistent with the process
accounting for at least part, and perhaps all, of the enhancement
apparently required in the helium resonance lines in the quiet Sun. 

A complementary study of the other of Jordan's
(1980) suggestions, enhanced collisional excitation by
suprathermal electrons of non-local origin, is presented in this paper. 
Radiative transfer codes usually assume that the electron velocity
distribution functions (EVDFs) are Maxwellian throughout the model
atmospheres, but observations of the solar wind (e.g.\ Scudder 1994)
and numerical studies of the form of the EVDF in the
solar transition region suggest that significant departures from
Maxwellian distributions may occur. This may result from some
acceleration process maintaining a non-Maxwellian distribution in the
low TR or chromosphere (Scudder 1992a; Vi\~nas, Wong \& Klimas 2000)
or from the transport of
high energy electrons from the upper TR and corona
\cite{es83,lb90}. In the second case at least, the influence of the
suprathermal electrons would depend on the temperature gradient. 

The presence of an enhanced suprathermal electron population would
simulate an increase in electron temperature in transitions with
excitation energies well above the local thermal energy, and hence
lead to larger excitation rates. 
Transitions with $W$ smaller than the energy at which the
suprathermal electron population becomes important would not be
affected significantly. 
Postulated non-Maxwellian EVDFs generally show departures from the
local Maxwellian distribution which increase monotonically with
energy, so that their effects on ionization rates can be even
larger than on line excitation. Essentially the same effect
limits the effectiveness of non-thermal transport of helium at
temperatures much higher than those of normal line formation. 
In the case of collisional ionization by non-local suprathermal
electrons the effect is a shift of the ionization equilibrium;
increased collisional ionization rates raise the population of each
ion at lower temperatures. Emission could therefore be increased
because it would occur over a larger, more dense region.

In a study of Si~{\sc iii} emission line ratios, Pinfield et al.\
\shortcite{pke99} claim to have observed both an enhancement of the
intensity of the line with the largest value of $W/kT$ and a reduction
in peak line formation temperature in the quiet Sun with respect to a
coronal hole. In light of the above discussion, this is evidence of 
excitation by an enhanced suprathermal population of electrons whose
importance depends on the temperature gradient in the transition region. 

Although the same mechanism has often been suggested to explain the
anomalous intensities of the helium resonance lines, few detailed
calculations of the possible effects on helium have been made,
perhaps because of the difficulty of solving the Boltzmann equation
for the EVDF and the equation of radiative transfer simultaneously. 
Shoub \shortcite{es83} computed non-Maxwellian collision rates for the
ionization of He~{\sc i} and He~{\sc ii} and the excitation of the
584.3-\AA\ and 303.8-\AA\ lines, but did not calculate line intensities
for comparison with observations. Anderson, Raymond \& Ballegooijen
\shortcite{arb96} computed intensities for the He~{\sc ii} 303.8-\AA\
line using non-Maxwellian EVDFs, but did not include radiative
transfer, and made no study of He~{\sc i}. 

In this paper I present the results of radiative transfer calculations
for cases of both Maxwellian and non-Maxwellian collisional
excitation. The former illustrate the problems found in attempting to
model the formation of the helium resonance lines; the latter are used
in a study of the observable effects that non-Maxwellian EVDFs
might have on He~{\sc i} and He~{\sc ii} lines. The calculations allow
an extensive comparison of the predictions of the enhanced excitation
mechanism with observations. 
The atomic and atmospheric models used in the radiative transfer
calculations are described in Section \ref{sec2}, and 
the results of radiative transfer calculations performed with
Maxwellian collision rates are presented in Section
\ref{sec3}. Section \ref{sec4} describes how the effects of non-local
electrons are simulated. The results of these latter calculations are
presented and discussed in Section \ref{sec5}, and the conclusions are
summarized in Section \ref{sec6}. 

\section{Atomic and atmospheric models}
\label{sec2}
The radiative transfer modelling of the solar helium spectrum
was carried out using version 2.2 of the {\sc multi}
code \cite{sc85a,mc86}. This code may be used to solve non-LTE
problems in semi-infinite plane-parallel one dimensional model
atmospheres. In order to calculate an emergent spectrum, the code
requires a model atom, a model atmosphere, and a set of abundances
(which are used to calculate background opacities - these are
generally computed assuming LTE, but in the calculations reported
here, the opacity due to hydrogen was computed in non-LTE). It also
allows the specification of the coronal radiation field as a boundary
condition at the upper edge of the model atmosphere.

\subsection{The model atom}
The atomic model used in this study comprises 29 bound levels of
He~{\sc i} ($1snl~^{1,3}L$ terms with principal quantum number $n \leq 5$),
six bound levels of He~{\sc ii} with $n \leq 6$, and the He~{\sc iii} ground
state. The atomic data used before modifications were made to
incorporate collisional excitation and ionization by non-Maxwellian
EVDFs are listed below.

The He~{\sc i} part of the model was initially based on the 30 level
He~{\sc i} model of Andretta \& Jones \shortcite{aj}, but using newer
and/or more complete data for some of the parameters, and was extended to
include He~{\sc ii}. Greater detail was included in the neutral stage,
in order to facilitate investigation of the relationships between
the 584.3-\AA\ and 537.0-\AA\ lines, which have been observed
extensively with {\em SOHO}. Previous work by Hearn \shortcite{ah69a}
and Andretta \& Jones (1997) suggested that the formation of these
lines depends on a combination of collisional and radiative
transitions between many levels, and is not necessarily dominated by
either direct collisional excitation or
photoionization-recombination (PR). It was therefore important to include
as many levels and processes as possible in this part of the model. 
In He~{\sc ii} most attention was paid to the 303.8-\AA\ resonance
line, which also appears in {\em SOHO} {\sc cds} spectra.

The energies of the He~{\sc i} levels are taken from Martin
\shortcite{wm87}, averaging over the fine structure in the
triplet levels. The fine structure in the triplet terms is not
included, which is a valid approximation in the rate equations since
the energy separation of the sub-levels is small enough for
collisional transitions between them to keep their populations in the
proportions of their statistical weights. As noted by Andretta \&
Jones (1997), although this assumption may not be valid in the
$2p$~$^{3}P^{o}$ term, where the separation of the $J$-states is
greatest, radiative transitions will also tend to populate the
sub-levels in the same proportions in the conditions encountered in
the atmospheric models used. 
The energies of the He~{\sc ii} levels are from Kisielius, Berrington
\& Norrington \shortcite{kbn96} for $n \leq 4$, from Bashkin \& Stoner
\shortcite{bs75} for $n \geq 5$. States of different $l$ were summed
over according to their statistical weights.  

Oscillator strengths for most of the allowed transitions in
He~{\sc i} are provided by Drake \shortcite{gd96}, while values for the few
transitions not covered there are taken from Theodosiou
\shortcite{ct87}. The rate for the spin-forbidden electric dipole
$1s2p$~$^{3}P_{1}$ -- $1s^{2}$~$^{1}S$ transition is taken from Drake \&
Dalgarno \shortcite{dd69}, and the two photon rate for the
$1s2s$~$^{1}S$ -- $1s^{2}$~$^{1}S$ transition is also included, from
Bassani \& Vignale \shortcite{bv82}. The He~{\sc ii} oscillator strengths are
taken to be hydrogenic, using the results of Wiese, Smith \& Glennon
\shortcite{wea66}. 

Stark broadening parameters for He~{\sc i} are from Dimitrijevi\'c \&
Sahal-Br\'echot (1984,1990) where available, with the remainder being
estimated using the formula of Freudenstein \& Cooper (1978). 
The Stark widths of Griem \shortcite{hg74} are used for He~{\sc ii}.
Van der Waals broadening is treated using the tables of Deridder \& Van
Rensbergen \shortcite{dvr76}, taking advantage of alterations made to
{\sc multi} by Rowe \shortcite{ar96}.

For most of the levels of He~{\sc i} photoionization cross-sections are
drawn from the Opacity Project database \cite{ms87}; they
were calculated by Fernley, Taylor \& Seaton \shortcite{fts87}. The
broadest, most prominent resonances in the cross-sections are
represented approximately in the model, particularly in the resonance
continuum, as there is some variation of the coronal radiation field
in the wavelength range of the resonances. The 4~$^{3}P$, 4~$^{1,3}F$,
5~$^{3}P$, 5~$^{1,3}F$, and 5~$^{1,3}G$ levels of He~{\sc i} are
treated as hydrogenic, as are the photoionization cross-sections for
He~{\sc ii} \cite{mp35}.

Photoionization is important in establishing the ionization
balance between all three stages of helium and is important in the
formation of some of its lines. The effects of coronal radiation
shortwards of 504 \AA\ are therefore included in the calculation of
ionization and recombination rates. The coronal illumination was
represented using the solar EUV irradiance model of Tobiska
(1991,1993). The spectrum of radiation at minimum coronal
activity was used to give incoming intensities at the solar surface
representative of the average quiet corona. Photoionization of He~{\sc
i} occurs directly from the ground state, by coronal radiation, and by
a two-stage process, in which He~{\sc i} triplet states, collisionally
excited from the ground, are rapidly photoionized (directly or by
successive photoexcitations followed by ionization) by the
photospheric radiation field. The importance of this two-stage process
was recognized in early calculations by Athay \& Johnson
\shortcite{aj60} and Hearn \shortcite{ah69a}, and by Andretta \&
Jones \shortcite{aj}, who emphasized how the singlet and triplet
systems of He~{\sc i} have quite different roles in the ionization
balance. Collisional excitation of singlet levels tends to lead more
often to decays back to the ground, to which the 
singlet levels are strongly connected by the allowed radiative
transitions $1snp$~$^{1}P^{o}$ -- $1s^{2}$~$^{1}S$. In the triplet
system, the metastable 2 $^{3}S$ level acts like a ground state,
and recombinations to higher triplet levels tend not to lead to cascade
decays, but more often result in re-ionization by the photospheric
radiation field. Recombinations to singlet states generally result in
radiative cascades to the ground. 

Collisonal excitation rates between bound levels of He~{\sc i} are
calculated, where possible, using the collision strengths of 
Lanzafame et al.\ \shortcite{lea93}, as
these are valid over a large temperature range. Collision rates in the
resonance series are among those calculated using these
data. Collision strengths for most of the remaining transitions in
He~{\sc i} are provided by Sawey \& Berrington \shortcite{sb93}, and
gaps are filled using data from Mihalas \& Stone \shortcite{ms68} and
Benson \& Kulander \shortcite{bk72}. Some of these early data are
expected only to be correct to an order of magnitude. 
Collision strengths for bound--bound transitions in He~{\sc ii} are
drawn from Aggarwal et al.\ \shortcite{aea92} for $n \leq 3$ and from 
Aggarwal, Berrington \& Pathak \shortcite{aea91} for transitions
involving $n = 4$. The remaining He~{\sc ii} collision strengths are
calculated using the expression given by Mihalas \& Stone (1968).  

Collisional ionization rates from all levels of He~{\sc i} and He~{\sc
ii} are calculated using expressions from Mihalas \& Stone
(1968). Older data have been used in preference to newer information
because the more recent work is not general enough. Bray et al.\
\shortcite{bea93}, for example, give the ionization cross-section for
only the ground state of He~{\sc ii}; the equivalent cross-section
provided by Mihalas \& Stone (1968) is in fact very similar, but they
also give expressions for ionization from excited states.

Dielectronic recombination to He~{\sc i}, for both Maxwellian and
non-Maxwellian electron distributions, was investigated in detail
(Smith 2000), but was found to have very little effect on the
formation of the 584.3-\AA\ and 537.0-\AA\ lines. 

\subsection{Model atmospheres}
\label{sec2.2}
\begin{figure}
\centering
\begin{minipage}{3.7in}
\epsfxsize=3.7in
\epsfbox{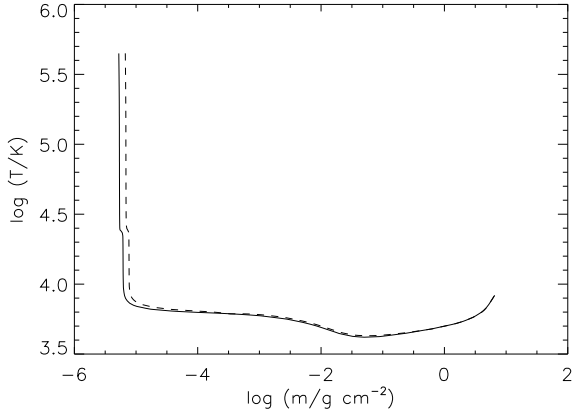}
\end{minipage}
\caption[VAL C and D model atmospheres.]{Temperature versus column
mass density ($m$) in the VAL C (solid) and D (dashed)
model atmospheres.\label{fig:2e}}
\end{figure}

Since the detailed physical processes responsible for heating the 
chromosphere, transition region, and corona remain unknown, most of
the atmospheric
models suitable for use in radiative transfer calculations are
semi-empirical. Their temperature structures are chosen to reproduce
observations of spectral lines and continua, rather than being derived
from the energy balance \emph{ab initio}. The VAL model atmospheres
(Vernazza, Avrett \& Loeser 1981, hereafter VAL), are successful in many
respects and are the most often used solar models of this type.
The models were constructed to fit six brightness components (A--F)
observed by the Harvard instruments on Skylab. The C component
represents the
average quiet Sun, and the VAL C model was initially adopted in the
present work. The VAL D model, representing the average quiet Sun
network, was also tested against observations. The run of temperature
against column mass density in both models is shown in Figure
\ref{fig:2e}. 

Fontenla et al.\ \shortcite{fal93} modified the VAL model atmospheres
in an effort to improve the physical basis of the
models. These new (FAL) models have semi-empirical chromospheric
structure based on the VAL models, but have transition regions derived
from energy balance considerations including the the effects of
ambipolar diffusion of H and He. These models produce results
consistent with observations of the hydrogen lines without need for
the plateau at about $2 \times 10^{4}$ K in the VAL models (see Figure
\ref{fig:2e}), but do not reproduce observed helium resonance line
intensities well. The FAL models neglect the effects of
turbulence (except in the equation of
hydrostatic equilibrium), and it is not clear to what extent the
steady state diffusion flows in the models are present in the dynamic
solar atmosphere. The models also do not include the effects of
diffusion on any elements heavier than helium, lines of which must be
compared with the helium lines in any explanation of the formation of
the latter. These factors suggest that the representation of the
transition region in the FAL models is not necessarily superior to
that in the semi-empirical VAL models. For these reasons, the VAL
models were used in preference to FAL in the present work. 

There are some problems with the VAL models. Calculations
using the VAL C and D model atmospheres produce 
intensities in the C~{\sc ii} lines around 1335 \AA\ about a factor of
six larger than the values observed by MJ99 \cite{grs00}. This appears
to be due to the influence
of the `Lyman $\alpha$ plateau' in the VAL models at about $2 \times
10^{4}$ K, which was included in order to put enough material in the
appropriate temperature range to reproduce the observed intensities of
lines of H and He, particularly the very important H
Ly$\alpha$ line. The plateau produces a local peak in the EMD in the
temperature range of the formation of both the He~{\sc i} resonance
series and the C~{\sc ii} lines. MJ99 found that this peak also
causes Si~{\sc iii} line intensities to be over-produced by a factor
of at least 2 with respect to their observations.
In view of this problem with the VAL models, and in order to make
comparisons with MJ99 more meaningful, new model atmospheres were
constructed to be consistent with emission measures derived from
observed intensities of TR lines of elements other than helium.

The new models were chosen to represent the network rather than cell
interior regions. Models of the magnetic field generally show flux
tubes confined to the network boundaries at photospheric levels
expanding in the chromosphere and TR to fill the atmosphere in the
corona (see e.g.\ Gabriel 1976). Such models suggest that network
regions may be more reasonably modelled with radial magnetic fields
than cell interiors. The effects of non-local electrons would be
expected to be greatest where the magnetic field extends directly into
the corona. Also, the VAL chromospheric model may not be valid in cell 
interiors, as the time-dependent models of Carlsson \& Stein
\shortcite{cs95} suggest that a classical chromosphere with an
outward-increasing temperature may not exist in non-magnetic
internetwork regions. 

The {\sc multi} code requires, as input for a (static) model atmosphere, the
run of mass column density, temperature, electron density, and microturbulent
velocity. Below $T\sim10^{4}$ K these figures were taken directly from
the VAL D network model. The transition region and coronal parts of
the models were developed from emission measure distributions
using methods described by Philippides \shortcite{dp96} and McMurry
\shortcite{am97}. The EMDs used to derive these parts of the new
models are described in detail in Paper I and only brief details are
repeated here. In the upper TR (above log~$T_{\textrm{e}}$ = 5.3), the EMDs
were derived assuming energy balance between radiation losses and the
divergence of the classical conductive flux from the corona. The coronal
emission measures were chosen to fit the EMD to MJ99's observations of
Mg~{\sc ix} and Mg~{\sc x}. As discussed in Paper I, this requires 
pressures higher than were observed by MJ99 in the network, but
similar to those in the VAL models. Below log~$T_{\textrm{e}}$ = 5.3,
the lower transition region of model S was derived directly from
MJ99's network EMD. As explained in Paper I, the form of the EMD is
uncertain, as the {\sc cds} 
and {\sc sumer} lines could not be observed at the same locations at
the same time. A second model was therefore constructed. Model X was made 
using the EMD derived from HST observations of the G8~V star $\xi$ Boo
A, scaled to the minimum of the MJ99 network EMD (see Paper I).
The lower transition region parts of the two models were derived
assuming hydrostatic equilibrium (including a contribution to the
pressure from turbulent motions) using methods discussed by Jordan \&
Brown \shortcite{jb81} and more recently by Harper
\shortcite{gh92}. 

\begin{figure}
\centering
\begin{minipage}{3.7in}
\epsfxsize=3.7in \epsfbox{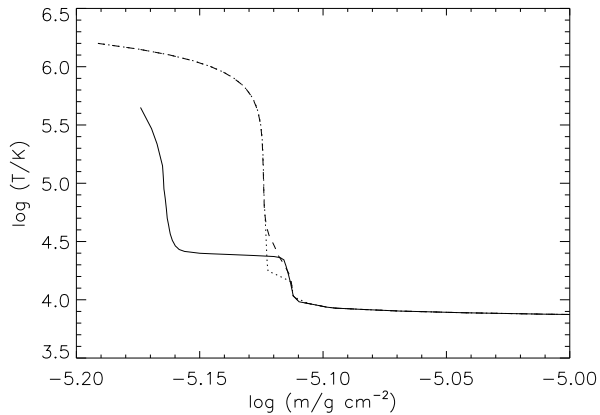}
\end{minipage}
\caption{Temperature versus column mass density in network
models S (dashed) and X (dotted) in the region where they differ from
the VAL D model (solid). \label{fig:3h}}
\end{figure}

The resulting transition region models were grafted separately on to
the VAL D model chromosphere. Running {\sc multi} in hydrostatic 
equilibrium with a 9 level hydrogen atom produced new values of
$x(T) = N_{\textrm{\sc h}}/N_{\textrm{e}}$ and of the height at the
base of the transition region, which were used to improve the transition
region parts of the models. Self-consistent models were found by
iterating this process to convergence in $x(T)$.
The resulting models are shown in Figure \ref{fig:3h}, which shows how
the temperature plateau in the VAL D model is suppressed in the new
network models. The full models used in the
calculations are given in an appendix in Tables \ref{tabA7} and
\ref{tabA8}.

The elemental abundances used in the calculations were 
the photospheric abundances of
Grevesse, Noels \& Sauval \shortcite{gea92} and Anders \& Grevesse
\shortcite{ag89}. The helium abundance is taken as
$N_{\textrm{{\sc h}e}}/N_{\textrm{\sc h}} =
0.098$ and is assumed to be constant through the
atmosphere. Calculations by Hansteen, Leer \& Holzer \shortcite{hlh97}
predict that the helium abundance could vary significantly
in the solar atmosphere. Their models show a \emph{decreased} helium
abundance in the transition region, and so cannot explain the
apparent enhancement of the helium line intensities. 

\section{Results using Maxwellian collision rates}
\label{sec3}
In this Section the results of calculations assuming purely Maxwellian
electron distributions are discussed, and compared with the 
observations of MJ99. Calculations were performed both
with and without the quiet coronal illumination described in Section
\ref{sec2} in order to investigate the importance of
photoionization-recombination (PR) in the formation of the helium
resonance lines. The discussion focusses on the He~{\sc ii} resonance
line at 303.8 \AA\ and the first two lines of the He~{\sc i} resonance
series, at 584.3 \AA\ and 537.0 \AA. The He~{\sc i} triplet lines
(e.g.\ 10830 \AA) are not discussed here; they were discussed in
detail by Andretta \& Jones (1997), and nothing has been found that
contradicts their analysis.

\begin{table}
\caption{Quiet Sun intensities and line ratios observed by Macpherson
and Jordan (1999). For averages over the network see also Jordan et
al.\ (2001).\label{tab1}}
\begin{center}
\begin{tabular}{ccccc}
\hline
 Region: & Strong & Typical & Cell & Average \\
 & network & network & interior & quiet Sun \\\hline
$\lambda$ & \multicolumn{4}{c}{Observed intensity} \\
(\AA) & \multicolumn{4}{c}{(erg cm$^{-2}$ s$^{-1}$ sr$^{-1}$)} \\\hline
584.3 & 613 & 613 & 380 & 487 \\
 & $\pm93.0$ & $\pm114$ & $\pm126$ & \\
537.0 & 72.8 & 70.9 & 44.7 & 56.9 \\
 & $\pm10.5$ & $\pm12.3$ & $\pm17.1$ & \\
303.8 & 10050 & 8289 & 5066 & 6654 \\
 & $\pm1249$ & $\pm1216$ & $\pm1540$ & \\\hline
 & \multicolumn{4}{c}{$I(\lambda)$/$I$(584.3 \AA)} \\\hline
537.0 & 0.119 & 0.116 & 0.118 & 0.117 \\
303.8 & 16.4 & 13.5 & 13.3 & 13.7 \\\hline
\end{tabular}
\end{center}
\end{table}

The intensities
observed by MJ99 in different regions of the quiet Sun are given in
Table \ref{tab1}, with an `average quiet Sun' intensity calculated for
comparison with results from the VAL C model. On the basis of a
comparison of the intensity classifications of MJ99 and VAL, this
average is weighted as 0.54 $\times$ `cell interior' intensity +
0.40 $\times$ `typical network' intensity + 0.06 $\times$ `strong
boundary' intensity. The intensities given by MJ99 were derived using
the calibration in the {\sc cds} software at the time of the
observations (1997), with the modifications suggested by Landi et al.\
(1997). Jordan et al.\ (2001) show how these intensities would be
changed by adopting the calibration derived by Brekke et al.\
(2000). When the most recent calibration is used, the main effect is a
reduction of the He~{\sc ii} 303.8-\AA\ line intensities by a factor
of 2.2. 

\subsection{Quiet coronal illumination}
The intensities of the resonance lines computed with quiet coronal
illumination are given in Table \ref{tab2}.  
As found by Andretta \& Jones (1997) in their study of the He~{\sc i}
lines, excitation of the 2~$^{1}P$
level occurs directly from the ground by collisions and, at least as
rapidly, through the excitation of levels with strong allowed radiative
transitions to 2~$^{1}P$. Over much of the region of line formation,
the most important populating process is by photoexcitation from
2~$^{1}S$, which itself is mainly excited by collisions from the
ground. Allowed transitions from other singlet states excited from the
ground, particularly 3~$^{1}S$ and 3~$^{1}D$, are also important;
net rates for the two-step processes are comparable to direct
collisional excitation of 2~$^{1}P$ from the ground. The contribution
function peaks at log~$T_{\textrm{e}} \simeq 4.45$. Collisional excitation
(direct and indirect) dominates above this temperature. At lower temperatures
($4.0 <$ log~$T_{\textrm{e}} < 4.4$) collisional excitation is less important
than radiative recombination cascades through the singlet levels
(3~$^{1}S$, 3~$^{1}D$ etc.). 

\begin{figure}
\centering
\begin{minipage}{4in}
\epsfxsize=3.6in
\epsfbox{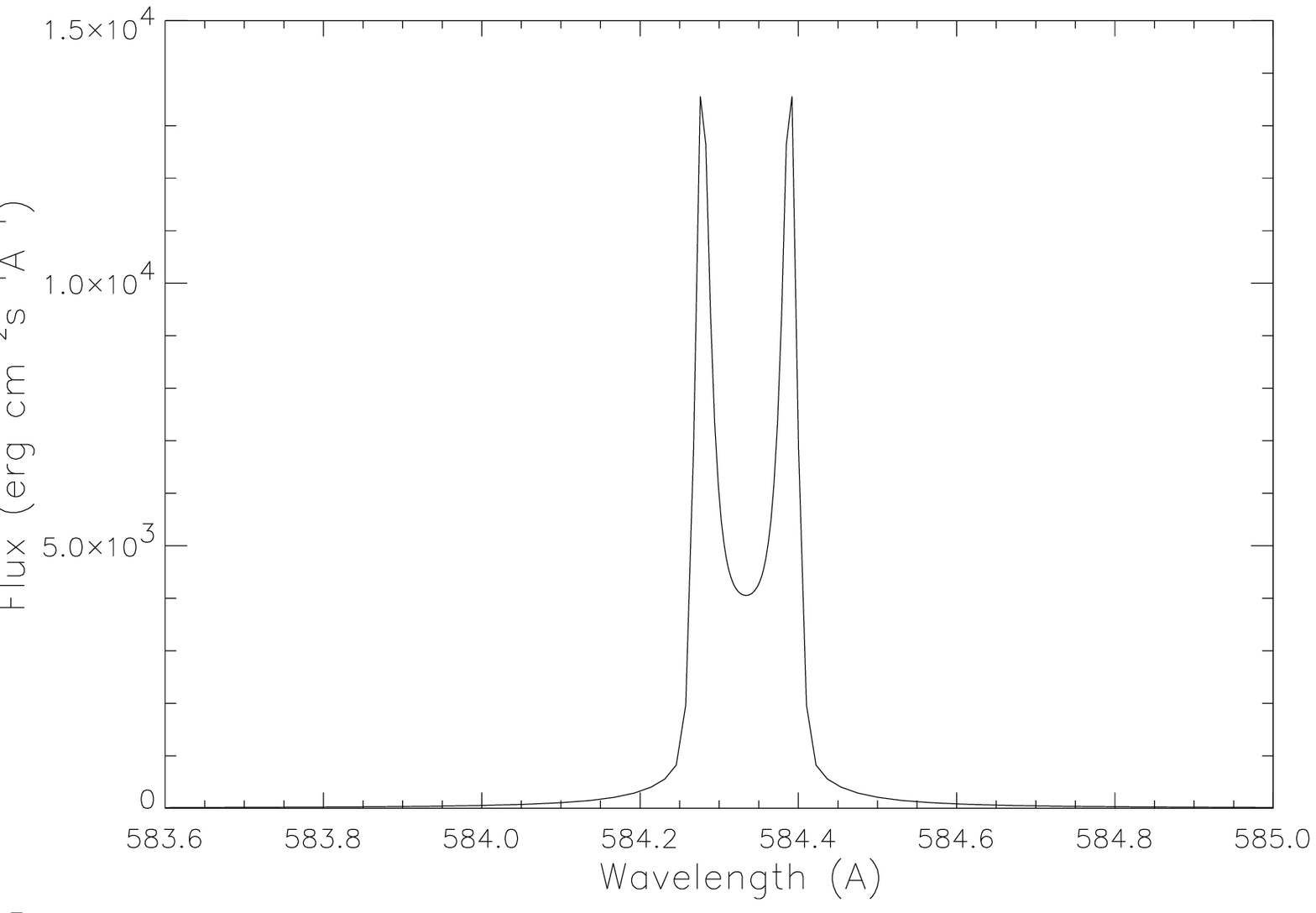}
\end{minipage}
\begin{minipage}{4in}
\epsfxsize=3.6in
\epsfbox{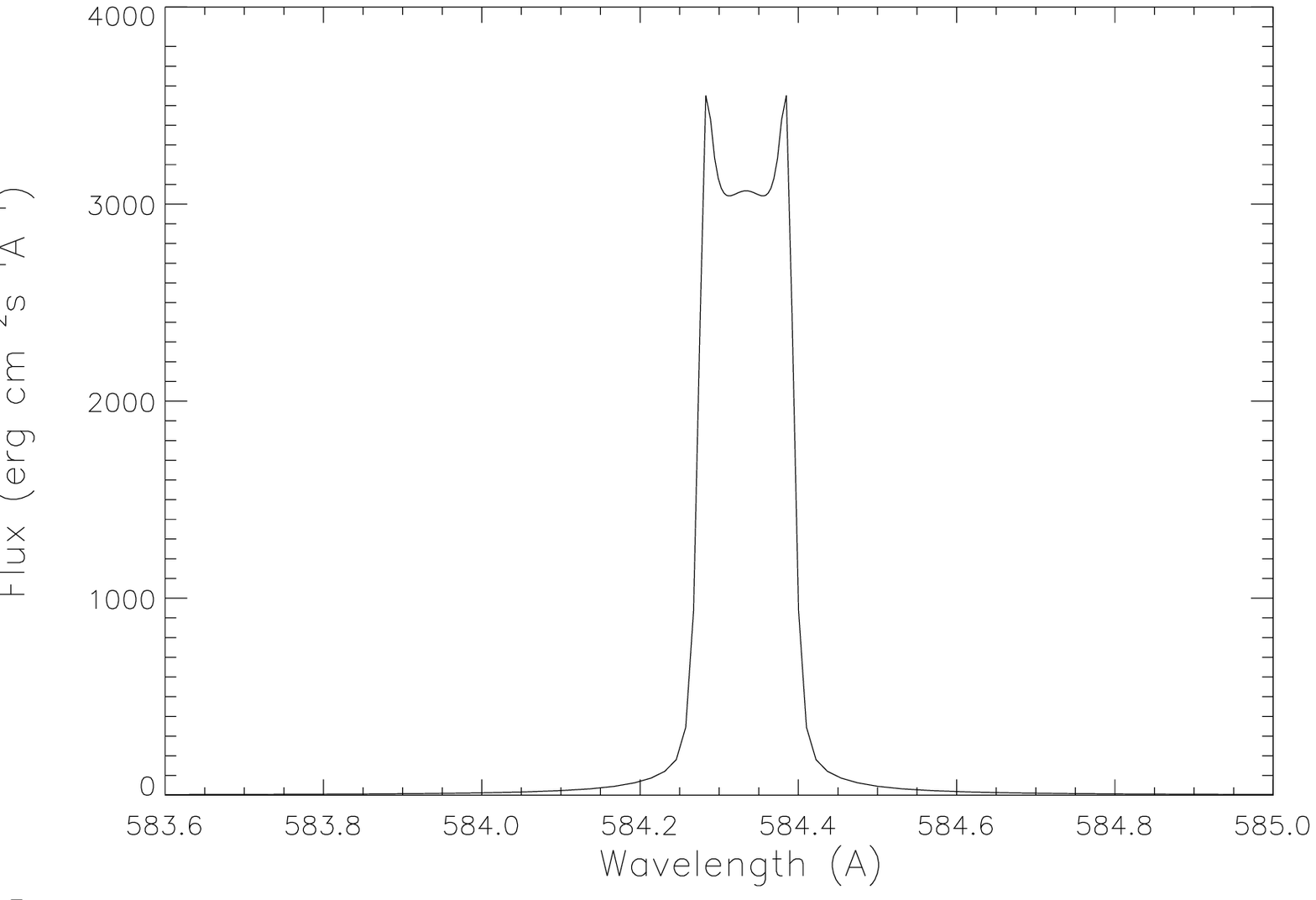}
\end{minipage} 
\caption{He~{\sc i} 584.3-\AA\ line profiles computed using the VAL C
(top) and network S (bottom) models with quiet coronal illumination. 
\label{fig1}}
\end{figure}

\begin{figure}
\centering
\begin{minipage}{4in}
\epsfxsize=3.2in
\epsfbox{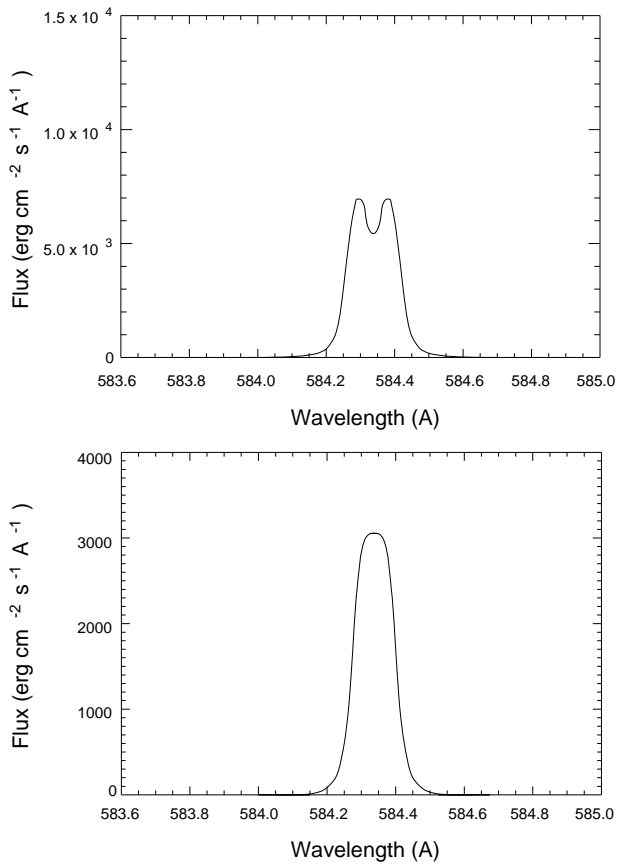}
\end{minipage}
\caption{He~{\sc i} 584.3-\AA\ line profiles computed using the VAL C
(top) and network S (bottom) models (shown in Figure \ref{fig1}),
convolved with a Gaussian function representative of the instrumental
broadening of the {\sc sumer} instrument. \label{fig2}} 
\end{figure}

Using the VAL C model atmosphere, the calculated intensity of the
584.3-\AA\ line is smaller (by about 20 per cent) than the quiet Sun
average of MJ99's observed intensities (see Tables \ref{tab1} and
\ref{tab2}). Output from the code suggests that 
the line is optically thick, with an optical depth at line centre
between 10 and 100, but it is \emph{effectively} thin, as found by Hearn
\shortcite{ah69a}; i.e.\ almost all photons
created in the line escape the atmosphere. The
computed line profile has a deep central self-reversal (see
Figure \ref{fig1}). 
At the spectral resolution of the {\sc sumer} instrument (about 0.04
\AA; Wilhelm et al. 1995), such a reversal would be obscured by
instrumental broadening. A convolution of the computed profile with a
Gaussian instrumental width of 0.04 \AA\
shows a much smaller self-reversal (see Figure \ref{fig2}). This
convolved profile is more consistent with observations of the quiet Sun.
Using {\sc sumer}, MJ99 found evidence for a small self-reversal in cell
interior regions, but none in the network. Peter \shortcite{hp99}
found the mean quiet Sun profile to be flat-topped. In coronal holes,
where the line has a larger optical depth, both Peter \shortcite{hp99}
and Jordan et al.\ (2001) observed self-reversed profiles.
The computed width of the
line (FWHM $\simeq$ 0.13 \AA) is similar to those obtained in
calculations by Andretta \& Jones \shortcite{aj}, Fontenla et al.\
\shortcite{fal93,fal02}. It also compares well
with the mean width of 0.14 \AA\ observed by Doschek, Behring \& Feldman
\shortcite{dbf74}, and the widths $\leq$ 0.13 \AA\ observed by
Peter \shortcite{hp99}. 

The 537.0-\AA\ line forms by similar processes to the 584.3-\AA\ line,
with contributions to the population of the upper level, 3~$^{1}P$,
from direct collisional excitation and radiative transitions from
other levels populated by a combination of collisional excitation and
recombination. The most important
intermediate levels in this case are 2~$^{1}S$, 4~$^{1}S$ and
4~$^{1}D$. Unlike in the 584.3-\AA\ line, direct recombination to
3~$^{1}P$ is more important than cascade recombination. The overall
contribution function is similar to that of the 584.3 \AA\ line, but
peaks at slightly lower temperatures (log~$T_{\textrm{e}} \simeq 4.40$).  
The 537.0-\AA\ line has an optical thickness smaller
than than that of the 584.3-\AA\ line, but is less likely to be
effectively thin, as explained by Hearn (1969). The optical
depth at line centre is of order 10 in calculations with the
VAL C model atmosphere. The computed intensity of the line is about
half the quiet Sun mean of MJ99's observations.

The formation of the He~{\sc ii} resonance line is dominated by direct
collisional excitation, its contribution function peaking at a
temperature of log~$T_{\textrm{e}} = 4.9$. There is, however, a small but not
negligible contribution to the population of the $n = 2$ level from
radiative cascades from higher levels. The levels with $n \geq 2$ are
similarly populated by collisional excitation at high temperatures and
recombination cascades at low temperatures. Recombination becomes
increasingly important as $n$ increases (see below for a brief
discussion of the 1640.4 \AA\ multiplet). The 303.8 \AA\ line is
marginally optically thick, with optical depth at line centre greater
than 1 but less than 10. This agrees with separate approximate
calculations, using the parameters of the radiative transfer
models, predicting optical depths in the He~{\sc ii} line of order 5
(and of order 100 in the He~{\sc i} 584.3-\AA\ line) . The line
is certainly effectively optically thin. In
results from the VAL C model, the computed line profile is
approximately Gaussian, and shows no sign of self-reversal (see Figure
\ref{figY} in Section \ref{sec5}). The line
width (FWHM $\simeq$ 0.08 \AA) is somewhat smaller than the value of
0.10 \AA\ observed by Doschek et al.\ \shortcite{dbf74} and Cushman
\& Rense \shortcite{cr78}, but is very similar to that computed by Fontenla
et al.\ (1993). Fontenla et el.\ \shortcite{fal02} predict greater widths if
significant flows are present. The integrated intensity calculated for
the VAL C model is a factor of more than 3 smaller than the quiet Sun
average of MJ99's observations (or only a factor of 1.5 smaller when the
most recent {\sc cds} calibration is used), but is about twice
that computed by Fontenla et al.\ (1993).

MJ99 suggested that the appearance of the network in the helium
resonance lines (broader and with lower contrast than in other lines
formed in the low TR) could be explained by their large optical
depths. If the network structures are thicker in the line of sight than
perpendicular to it, photons would more likely escape from the edges
of the network into cell interiors. If then resonantly scattered back
into the line of sight, this would apparently broaden the network, and
might explain why the helium lines appear to be most enhanced with
respect to other TR lines in the cell interiors. Both the He~{\sc i}
and He~{\sc ii} resonance lines were found to be optically thick in
the calculations described here, but as these calculations were
performed in the plane-parallel approximation, detailed conclusions
about scattering from the network cannot yet be drawn, but await 
radiative transfer calculations with two-component atmospheric models.

Table \ref{tab2} also gives the intensities of the helium lines
computed using the VAL D network model and the new network models S
and X. The 584.3-\AA\ line intensity computed using the VAL D model is
close to the observed mean network value, but the ratio
$I$(537.0~\AA)/$I$(584.3~\AA) is again smaller than observed. The
computed 303.8-\AA\ line intensity is only 25 per cent smaller than was
observed by MJ99, but adopting the most recent calibration, the
computed intensity is \emph{larger} than
observed by a factor of 1.7. The VAL D model also predicts other low
TR lines (e.g.\ of C~{\sc ii}) to be stronger than observed by
significant factors. Models S and X were constructed to be more
consistent with the observed intensities of TR lines other than those of
helium. Significant changes in the lower atmospheric structure with 
respect to the VAL models lead to changes in the relative importance
of the various processes responsible for the formation of the He~{\sc i}
lines. Although the electron
pressures are higher in the new network models than in the
VAL C model, the intensities of the He~{\sc i} lines are smaller owing
to the absence of the temperature plateau which is present in the VAL
model. This reduces the amount of material present in the new models in the
temperature range where collisional excitation (direct and indirect)
is effective, which is limited from below by the temperature
dependence of the excitation rates and from above by the ionization of
He~{\sc i}. Optical depth at the 584.3-\AA\ line centre is still
greater than 10. PR provides a relatively greater contribution to the total
excitation rate than in the VAL calculations, particularly in model X. 

\begin{table}
\caption{Integrated intensities computed for the VAL C and D and network S
and X models with quiet coronal illumination.\label{tab2}}
\begin{center}
\begin{tabular}{ccccc}
\hline
Model atmosphere: & VAL C & VAL D & S & X \\\hline
$\lambda$ & \multicolumn{4}{c}{Computed intensity} \\ 
(\AA) & \multicolumn{4}{c}{(erg cm$^{-2}$ s$^{-1}$ sr$^{-1}$)} \\\hline
584.3 & 406 & 624 & 146 & 81.2 \\
537.0 & 31.2 & 50.7 & 17.1 & 6.78 \\
303.8 & 2020 & 6400 & 1163 & 988 \\\hline
 & \multicolumn{4}{c}{$I(\lambda)$/$I$(584.3 \AA)} \\\hline
537.0 & 0.077 & 0.081 & 0.117 & 0.083 \\
303.8 & 4.98 & 10.3 & 7.97 & 12.2 \\\hline
\end{tabular}
\end{center}
\end{table}

The intensities computed for the 584.3-\AA\ and 537.0-\AA\ lines using
model S are a factor of about 4 smaller than observed network
intensities, but the ratio of the computed intensities is close to the
mean quiet Sun network value of $0.116 \pm 0.015$ found by MJ99. That
value assumed the Landi et al.\ (1997) calibration of {\sc cds}; using
instead the most recent calibration 
reduces the ratio to $0.105 \pm 0.014$. While this is matched less well 
by the computed ratio, model S still produces a better fit than the
other model atmospheres. The computed profile of the 584.3-\AA\ line
is also quite close to that observed, showing only a small
self-reversal that is completely obscured by instrumental broadening
of 0.04 \AA\ (see Figures \ref{fig1} and \ref{fig2}). The intensities
computed using model X are smaller than observed by 
factors of 7.5 for the 584.3-\AA\ line and 10 for the 537.0-\AA\ line,
owing to the lower emission measure derived from the scaled $\xi$ Boo A
distribution used in the lower TR. The ratio of the line intensities
is smaller than observed. 

The value of the intensity ratio $I$(537.0~\AA)/$I$(584.3~\AA)
found using the VAL C model atmosphere is very similar to that
obtained by Andretta \& Jones \shortcite{aj} in an equivalent
calculation; they found the ratio to be larger in models in which the
VAL C plateau was removed. Similarly, in models S 
and X, where the plateau is suppressed, the value of the computed
ratio is larger than in results from the VAL models.  
The reduction in the size of the line forming region lowers the
optical depth in the 537.0-\AA\ line (The 584.3-\AA\ line is effectively
thin in all the atmospheric models tested here, but the 537.0-\AA\ line
is not effectively thin in the VAL C calculations). Thus, although
fewer photons are created in the 537.0-\AA\ line in the network models,
a greater proportion escape, allowing the intensity ratio
$I$(537.0~\AA)/$I$(584.3~\AA) to increase compared with the VAL C
case. The ratio is smaller in results from model X than in those from model
S, although the region where collisional excitation of the lines is
important is narrower in model X than in model S. In model X, the line
formation has a relatively greater contribution from PR from the thicker
region below $T_{\textrm{e}} \simeq 1.8 \times 10^{4}$ K, where the
optical depth in the 537.0-\AA\ line is greater. In recombination
cascades, levels linked to both 2~$^{1}P$ and 3~$^{1}P$ by allowed
radiative transitions generally decay to the former more rapidly.

In the region of formation of the He~{\sc ii} resonance line the new
models have electron densities smaller than in the VAL D
model, but have steeper temperature gradients, reducing the extent of
the emitting region and thus the emergent intensity. The line still
has optical depth greater than 1. The line
intensities computed using models S and X are factors of 7 and 8
smaller than the mean observed network value respectively. These
factors are reduced to 3--4 when the most recent {\sc cds}
calibration is assumed. The temperature gradient is slightly larger
and the density smaller in model X than in model S, explaining the
lower emission in model X. The ratio $I$(303.8~\AA)/$I$(584.3~\AA)
found using model X (12.2) is the closest of those computed to the
observed network ratio of 13.5. using the most recent {\sc
cds} calibration, the observed ratio is about a factor of 2
smaller, and the ratio found using model S is a closer match.
The computed profiles of the 303.8 \AA\ line, with
FWHM $\sim 0.1$ \AA, are broader than in the VAL C results because of 
higher turbulent velocities in the regions of line formation in the
new models.

The separate components of the He~{\sc ii} Balmer $\alpha$
multiplet at 1640.4 \AA\ are not included explicitly in the current
atomic model, but the total predicted intensity may be compared with
observations. In calculations using the VAL
C model with quiet coronal illumination recombination is more
important than collisional excitation in the formation of the
multiplet, and the computed intensity is almost a factor of
6 smaller than the value of 88 erg cm$^{-2}$ s$^{-1}$ sr$^{-1}$
observed in the quiet Sun by Kohl \shortcite{jk77}. Models S and X produce
even poorer matches to observations, predicting intensities at least an
order of magnitude too small. Wahlstr\o m \& Carlsson
\shortcite{wc94} found a slightly smaller discrepancy than in the 
results here, of a factor of about 4, but this is still a significant
problem that requires explanation. 

\subsection{Variations in coronal illumination}
Andretta \& Jones (1997) investigated the effects on the He~{\sc i}
lines of varying the intensity of the coronal radiation field by
multiplying the assumed `quiet' illumination by a constant factor at
all wavelengths. They
found that both the 584.3-\AA\ and 537.0-\AA\ lines increased in
intensity with increasing coronal illumination owing to increased
contributions to line formation from PR. The results found here using the
VAL C model agree with those of Andretta \& Jones (1997), and
calculations using models S and X show trends similar 
to results from their `no plateau' models. 
The effects on He~{\sc i} of increasing the coronal intensity
are therefore not discussed in detail (but are given in Smith
2000). Andretta \& Jones (1997) did not, however, study lines of
He~{\sc ii}.

Somewhat surprisingly, the He~{\sc ii} 303.8-\AA\ line, whose
formation in all cases is dominated by collisional excitation,
generally shows a decrease in intensity as the coronal radiation is
increased. This appears to be due to a shift of the peak He~{\sc ii}
ionization fraction to lower temperatures, reducing the collisional
excitation rate relative to the quiet corona case. This outweighs any
increase in the small contribution from recombination to the line,
although when the coronal irradiance is reduced to zero in model X the
computed intensity of the He~{\sc ii} line is also slightly reduced.
Given that spatial variations of the intensities of the He~{\sc i} and
He~{\sc ii} resonance lines are very similar (e.g.\ MJ99), and
certainly not anti-correlated, this result supports the contention
that PR by coronal radiation does not dominate the formation of both
lines. 

The model atmospheres considered in this work are not appropriate to
coronal hole regions, where the density is a factor of 2 or more
smaller than in the quiet Sun, and the temperature gradient is much
shallower. Removing the coronal illumination included in the
radiative transfer calculations does however indicate how the lower
photoionizing flux in coronal holes might affect 
the helium resonance lines. The intensities of the
lines are observed to be a factor of 1.5 -- 2 smaller in coronal holes
than in the quiet Sun \cite{hp99,jea01}. Much smaller reductions in
the intensities of the He~{\sc i} 
lines are computed when the coronal radiation incident on the top of the
VAL C model is removed. In results from models S and X, where PR is
more important in forming the He~{\sc i} lines, more significant
effects are seen. The computed intensity of the 584.3 \AA\ line in
model S is reduced by a factor of 1.3 when the quiet coronal
illumination is removed, while with model X this causes the line
intensity to fall by a factor of almost 4. The observed increase in
the value of the He~{\sc i} line ratio $I$(537.0~\AA)/$I$(584.3~\AA)
in coronal holes is understandable in terms of the reduction of coronal
illumination. Such a trend is seen in the results from all of the
models, and the value of 0.134 found from model S with zero coronal
illumination agrees quite well with the mean value of
$0.130 \pm 0.014$ found in coronal hole observations by Jordan et al.\
(2001), although this observed value is decreased to 0.117 when the
most recent {\sc cds} calibration is adopted. 

In none of the models tested here is the He~{\sc ii}
resonance line reduced significantly in intensity by the removal of
coronal illumination. As this line is formed mainly by collisional
excitation its intensity would be expected to depend to a greater
extent on changes in the density and temperature gradient in coronal
holes than on changes in the coronal illumination. 
Given that the He~{\sc i} and He~{\sc ii} resonance lines respond
differently to significant changes in coronal illumination, the
observation that the ratio of the lines changes little between coronal
holes and the quiet Sun \cite{jea01} again implies that this is not the
factor controlling the changes in absolute intensity.

\section{Non-Maxwellian collision rates}
\label{sec4}
The calculations reported above show that none of the model
atmospheres tested gives a good match to all aspects of the
observations. Collisional
excitation by non-Maxwellian EVDFs was investigated as a possible
explanation of these discrepancies. Modifications were made to the
radiative transfer code to simulate the excitation and ionization of
the helium atom by EVDFs with enhanced suprathermal tails of a form
that has been suggested to exist in the solar transition region. 

\subsection{Transition region electron distributions}
In order to calculate the effects on collisional excitation and
ionization rates of a given non-Maxwellian EVDF, one needs to know its
shape as a function of
temperature. Strictly, an electron temperature cannot be defined for a
non-Maxwellian EVDF. In the distributions postulated to exist in the
transition region, however, the bulk of the electrons have a nearly
Maxwellian velocity spectrum, which is used to define the temperature. 

In any plasma, the EVDF is a solution to the Boltzmann (or
Fokker--Planck) equation:
\begin{equation}
\label{eq:4x}
\frac{\partial f}{\partial t} + \textrm{\boldmath $v$}. 
\frac{\partial f}{\partial \textrm{\boldmath $x$}} +
\frac{\textrm{\boldmath $F$}}{m}. 
\frac{\partial f}{\partial \textrm{\boldmath $v$}} =
\left(\frac{\partial f}{\partial t}\right)_{\textrm{coll}}
\end{equation}
where $f$ is the EVDF, {\boldmath $v$} and {\boldmath $x$} are electron
velocity and position, {\boldmath $F$} is the total force acting on the
electron, and the term on the right hand side is the rate of change of
$f$ with time due to collisional redistribution of electrons in velocity
space. The Maxwellian distribution is the homogeneous, steady state
solution found in a gas in thermodynamic equilibrium, which is
independent of {\boldmath $x$} and $t$.

Shoub (1983) showed that, although the solar TR is weakly
inhomogeneous (the mean free path of thermal electrons is at all
points small relative to the temperature and density gradient scale
lengths), the Spitzer--H\"arm \shortcite{sh53} solution is invalid at
high velocities, where the mean free path of an electron increases as
the fourth power of its velocity. Studies of the transition region
EVDF in the case
where the Knudsen parameter (the ratio of the electron mean free path
to the scale length) is large have been made in one of two
approximations. 

Some more recent work has focused on the process of `velocity
filtration,' proposed by Scudder (1992a,b). This postulates a
non-Maxwellian EVDF with an
over-population of suprathermal electrons at the base of the
transition region, which have large enough velocities to climb the
gravitational gradient into the corona, there forming the bulk of the
plasma at $T \sim 10^{6}$ K. Studies of this process assume
that the Knudsen parameter is of order one or greater, so that the
electron fluid is effectively collisionless, and the RHS of
equation (\ref{eq:4x}) is zero. Anderson et al.\ \shortcite{arb96} 
found that collisionless velocity filtration is inconsistent with
observations of transition region lines, and suggested that the
collision term in the Boltzmann equation should not be neglected.

In a different context, others (e.g Shoub 1983) have
considered a high velocity form of the Boltzmann equation including a
collision term due to Landau \shortcite{ll36} describing the
interaction of particles under inverse-square Coulomb forces. 
Solutions are computed for a model transition
region with a prescribed temperature profile derived semi-empirically
from observations or energy balance arguments. Such calculations have
been performed both for plane-parallel model TRs \cite{es83,lb90} and
for coronal loop models (Ljepojevic \& MacNeice 1988,1989). The
heating of the corona is assumed to occur by some unspecified
mechanism, and the coronal temperature is used as a boundary condition.

Whereas in the velocity filtration models a non-Maxwellian EVDF is
\emph{assumed} to exist as a boundary or initial condition, in the 
collisional models non-Maxwellian EVDFs are not assumed to exist
\emph{a priori}, but their existence and form in the transition region
are \emph{derived} as results of the steep temperature gradient. 
Given that the existence of a steep temperature gradient in the TR is
confirmed by many observations, whereas independent processes by
which non-Maxwellian EVDFs may be maintained are at present only
postulated, the collisional calculations seem to rest on more solid
assumptions. For this reason, and because Anderson et al.\ (1996) cast
doubt on the collisionless approach, the present work is based on
calculations of the EVDF using the Landau equation. 

Numerical solutions of the Landau equation in the TR result in
EVDFs, averaged over pitch angle, resembling (very slightly
underpopulated) local Maxwellians at low velocities, with more heavily
populated tails diverging from the Maxwellian at a few times the
thermal velocity. The
high velocity tail of the distribution at a point in the
lower TR is almost wholly populated by electrons originating from the
near-thermal parts of distributions present at higher
temperatures. The temperature gradient provides an excess of high energy
electrons moving downwards; as their speed-dependent mean free path
varies as $v^4$, these electrons are more influential than the similar
excess of low energy electrons moving up the gradient.

Shoub \shortcite{es82} obtained analytical solutions to the Boltzmann
equation using a linearized form of the Bhatnagar--Gross--Krook
\shortcite{bgk54} (BGK) model for the collision term. Using
this method, he found that the angle-averaged distribution functions
obtained were in reasonable agreement with those found in his
numerical work, with very similar enhanced suprathermal tails. 
The analytical solutions for the angle-averaged distribution in the
lower TR can be accurately approximated by a power law over a wide
range of suprathermal velocities. 

The distribution tested by Anderson et al.\ (1996) which most closely
approximated the results of Shoub (1982,1983) was the
$\kappa$ distribution (as suggested by Scudder 1992b). Anderson (1994)
suggested the use of a Maxwellian
EVDF with a power law tail attached above some critical speed (as
introducing a collision term makes the low speed part of the EVDF more
nearly Maxwellian). This is actually a much better
approximation to the collisional results of Shoub (1982,1983)
and Ljepojevic \& Burgess \shortcite{lb90}, in which the transition
from the Maxwellian form to a power law tail is generally sharper than
in a $\kappa$ distribution. For these reasons I have
chosen to use this Maxwellian plus power law EVDF in my calculations of
enhanced collision rates. The calculations are not formally
self-consistent, but assume that the EVDF throughout the model
atmospheres used in the radiative transfer calculations is of this
form when calculating the collisional excitation and ionization rates
for the model helium atom. 

\subsection{Parametrization of the EVDF}
\label{secA} 
The way in which the EVDF is parameterized in the present work was
influenced by the analytical BGK calculations made by Shoub
\shortcite{es82}. I take the angle-averaged EVDF to be locally Maxwellian
below a velocity $\xi_{\textrm{tail}}$, with a power law decline at higher
$\xi$, where $\xi$ is a dimensionless velocity $v/v_{\textrm{th}}$, and
$v_{\textrm{th}}$ is the thermal velocity $\sqrt{2kT/m}$. A feature of Shoub's
(1982) solutions which is common to similar work is that the value of
$\xi$ marking the departure from near-Maxwellian is only very weakly
dependent on the temperature defined by the Maxwellian bulk of the
electrons, so as a first approximation I take $\xi_{\textrm{tail}}$ to be
constant with respect to $T$. 

The slope of the power law tail found by Shoub (1982) was determined
by his choice of model atmosphere. This was an isobaric slab of
thickness $L$ in which energy is the transferred by (classical)
thermal conduction is constant. The corresponding temperature profile
takes the form 
\begin{equation}
\label{eq:4d}
T(z)^{7/2} = T_{\textrm{c}}^{7/2} + (T_{\textrm{h}}^{7/2} -
T_{\textrm{c}}^{7/2})z/L 
\end{equation}
where $T_{\textrm{c}}$ and $T_{\textrm{h}}$ are temperatures characterizing
the incoming EVDFs at the lower and upper boundaries of the atmosphere
respectively. This choice of atmosphere is an obvious
simplification. In the upper TR very little heating is required to
balance local radiation losses, and this can be provided easily by the
net conductive flux. In the lower TR, however, this energy balance
condition is not consistent with the rise in the emission measure
distribution below $T \sim 10^{5}$ K. Although the model atmosphere
Shoub (1983) used in his numerical work allows for radiative power
loss, the models used in both papers represent the low TR poorly, having
steeper temperature gradients than suggested by semi-empirical
models. This failing may not have a major effect on the accuracy of
the derived EVDFs in the lower TR, as the form of the
suprathermal tail is more dependent on the temperature structure in
the upper TR. In his analytical solutions, Shoub \shortcite{es82}
found that at $T = 2 \times 10^{4}$ K, electrons with $\xi \geq 5$
come mainly from $T > 10^{5}$ K, and at $T = 8 \times 10^{4}$ K,
electrons with $\xi \geq 4$ come from $T > 2 \times 10^{5}$ K.

\begin{figure}
\centering
\begin{minipage}{3.5in}
\epsfxsize=3.1in
\epsfbox{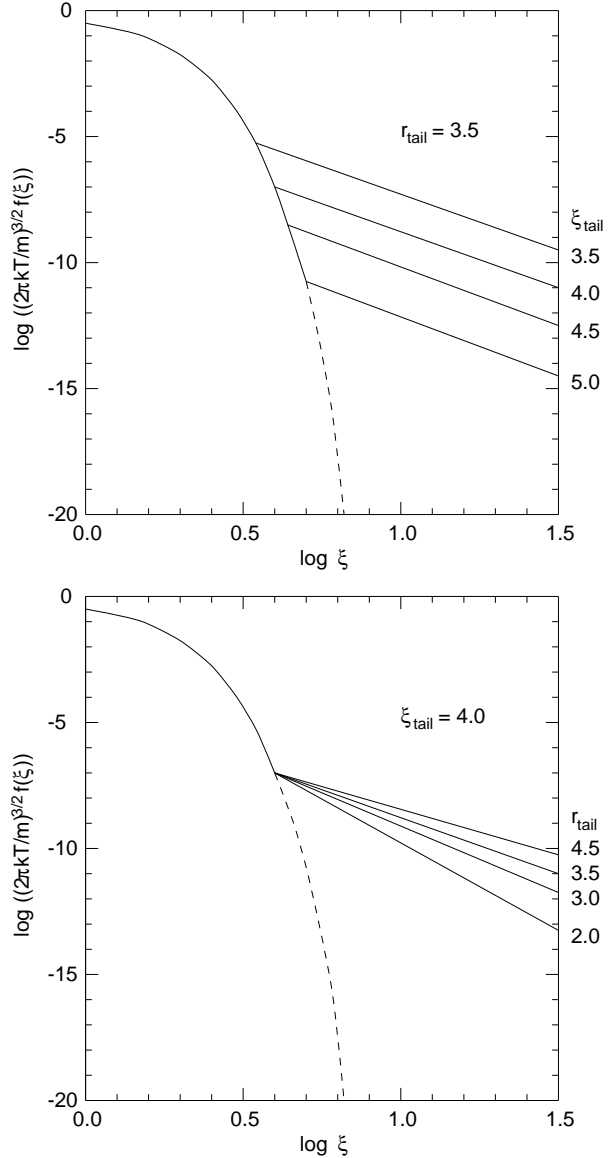}
\end{minipage}
\caption{Dimensionless angle-averaged EVDFs used to calculate enhanced
collisional excitation rates in helium. $(2\pi kT/m)^{3/2}f(\xi)$ is
given as a function of $\xi$. Plotted in this way, the shape 
of the EVDF is independent of $T$. Dashed lines show Maxwellian
EVDFs. The ranges of $\xi_{\textrm{tail}}$ and $r_{\textrm{tail}}$
used in the power law tail are illustrated.\label{fig:4a}}
\end{figure}

Shoub's (1982) assumptions resulted in a power
law decline in the tail of the EVDF varying as $v^{-31/7}$ (or
$\xi^{-31/7}$). He generalized his results, however, to allow for a
variation of the exponent $r = 7/2$ in equation (\ref{eq:4d}). The
corresponding index in the angle-averaged EVDF is $-(12/r + 1)$. This
notation has been used in the power law tails in the EVDFs used
here. A smaller value of $r$, here designated $r_{\textrm{tail}}$,
corresponds to a shallower temperature gradient low down and a steeper
one high up, and to a steeper power law in the tail of the EVDF. As
stated, Shoub's (1982) results have $r_{\textrm{tail}} = 3.5$ and 
are best represented in the formulation used here by taking $3.5 <
\xi_{\textrm{tail}} < 4.0$. Both $r_{\textrm{tail}}$ and
$\xi_{\textrm{tail}}$ are taken to be free 
parameters in the calculations reported here, and ranges of values
were investigated (see Figure \ref{fig:4a}). The parameter space was
defined by the results of Shoub (1982,1983) and Ljepojevic
\& Burgess \shortcite{lb90}. Calculations by the latter using two
different model atmospheres produced EVDFs in the lower TR that may be
approximated by taking $4.0 \leq \xi_{\textrm{tail}} \leq 5.0$ and $2.0 \leq
r_{\textrm{tail}} \leq 4.0$. Their solutions again showed $\xi_{tail}$ to be
almost independent of temperature for a given model, except at low
temperatures ($T \sim 2 \times 10^{4}$ K). They found
$\xi_{\textrm{tail}}$ to increase in this region, suggesting that the
enhanced suprathermal tail may not persist right down to the bottom of
the TR in models with more realistic (i.e. shallower) temperature
gradients in the lower TR than those used by Shoub (1982,1983). This
points to a potential problem with the formulation used here, which
assumes a power law tail to exist at all temperatures.

The low values of $\xi_{\textrm{tail}}$ required to represent Shoub's
(1982,1983) results are directly related to the steep
temperature gradient in the low TR of the model atmosphere he
used. Comparison of the atmospheric models used in the radiative
transfer calculations presented here (VAL C, models S, X) with
those used by Shoub (1982,1983) and Ljepojevic \& Burgess
\shortcite{lb90} suggest that values of $\xi_{\textrm{tail}} \geq
4.0$ are more appropriate to realistic models. Given this argument, a
parameter space of $3.5 \leq \xi_{\textrm{tail}} \leq 5.0$, $2.0 \leq
r_{\textrm{tail}} \leq 4.5$ was explored in these calculations, with
results for $\xi_{\textrm{tail}} \geq 4.0$ given greater credence.

\subsection{Collisional excitation and ionization rates}
In order to treat excitation and ionization by the non-Maxwellian EVDFs
described above in the radiative transfer calculations, options were
written into the {\sc multi} code to compute the relevant rates.
The collisional excitation and ionization rate coefficients for helium
were calculated by integrating electron collision cross-sections over
the speed and the electron speed distribution. If the collisional
excitation rate from level $i$ to level $j$ is $R_{ij} =
N_{\textrm{e}}C_{ij}$, the rate coefficient $C_{ij}$ is given by
\begin{equation}
C_{ij} = \int^{\infty}_{v(E_{ij})} 4\pi v^{3}f(v)\sigma(v)\textrm{d}v
\end{equation}
where $f(v)$ is the angle-averaged EVDF, $\sigma(v)$ is
the speed-dependent collision cross-section for the transition, and
$E_{ij}$ is the threshold energy of the transition.

For an EVDF consisting of a Maxwellian distribution with temperature $T$
for velocities smaller than $v_{\textrm{t}}$, with a power law decline of
$v^{-\rho}$ at higher velocities, $C_{ij}$ becomes 
\begin{eqnarray}
\label{eq4a}
C_{ij} &=& 4\pi \left( \frac{m}{2\pi kT} \right)^{3/2} \left\{
\int^{v_{\textrm{\sevensize{t}}}}_{v(E_{ij})} v^{3} \textrm{ exp} \left(
-\frac{mv^{2}}{2kT} \right) \sigma(v)\textrm{d}v \right.\nonumber\\
 & & \left.+\textrm{ exp} \left
( -\frac{mv_{\textrm{t}}^{2}}{2kT} \right)
\int^{\infty}_{v_{\textrm{\sevensize{t}}}} v^{3} \left
( \frac{v}{v_{\textrm{t}}} \right)^{-\rho} \sigma(v)\textrm{d}v \right\}
\end{eqnarray}
where the identities $v_{\textrm{t}} =
\xi_{\textrm{tail}}v_{\textrm{th}}$ and $\rho =
(12/r_{\textrm{tail}}+1)$ have been used to simplify the expression. If, at
temperature $T$, $v_{\textrm{t}} < v(E_{ij})$ (i.e.\ the
power law tail begins at an energy below threshold), then the first
integral vanishes and the lower bound on the second integral
becomes $v(E_{ij})$.

Analytical expressions for the collision cross-sections were used as
the cross-sections were needed up to large electron energies, where
few numerical data exist for many of the transitions. Keeping the
expressions for the rate coefficients in an analytical form also made
it simple to make $\xi_{\textrm{tail}}$ and $r_{\textrm{tail}}$ free
parameters, and
kept the expressions transparent in their dependence on these
parameters and on the temperature. The drawback of this approach is
that for some transitions newer and better numerical data exist which
are used in the standard atomic model described in Section \ref{sec2},
but not in these calculations of the
rates due to non-local electrons. The approach taken by Anderson et
al.\ \shortcite{arb96}, in which the EVDF at each height is written as
a weighted sum of Maxwellians, allowing use of numerical data
tabulated for Maxwellian distributions, is an improvement in this
respect, but the scarcity of high temperature data (for He~{\sc i} in
particular) is a problem for that method.

The specific expressions for the rate coefficients depend on the
cross-section used. For
collisional ionization of He~{\sc i} and He~{\sc ii}, expressions for
the cross-sections given by Mihalas \& Stone \shortcite{ms68} were
used. Mihalas \& Stone (1968) also provide bound-bound
collision cross-sections for He~{\sc ii}, using an easily-integrated
semi-empirical formula due to Hinnov \shortcite{eh66}, and for the
optically allowed transitions in He~{\sc i}. For the forbidden
transitions in He~{\sc i}, the expression due to Green
\shortcite{ag66} given by Benson \& Kulander \shortcite{bk72} was
used. Benson \& Kulander \shortcite{bk72} made order of magnitude
estimates for $f$ values used in the expression. I have made new
estimates by matching as closely as possible the collision strengths
given for $T_{\textrm{e}} = 6\times10^{3}$ K and $T_{\textrm{e}} =
2.5\times10^{4}$ K to more recently calculated 
numerical data used in the standard helium model \cite{lea93,sb93}.
In cases where the integration above $v_{\textrm{t}}$ could
not be performed analytically, it was performed
numerically; such integrals are estimated to be correct to about 5 per
cent. 

The modifications use collision rate data which differ from those in
original helium model described in Section \ref{sec2}. When the
modified rates are used with a very high value of $\xi_{\textrm{tail}}$ to
approximate the Maxwellian regime, this results in the computed
intensities of the He~{\sc i} resonance series and the He~{\sc ii}
303.8-\AA\ line differing from those calculated using the original
model by at most 10 per cent. Owing to approximations made for the
forbidden transitions of He~{\sc i} the non-Maxwellian
modifications cannot be used to investigate possible
non-Maxwellian effects on the triplet lines of He~{\sc i}. 

Enhanced collision rates were not adopted for all transitions in
He~{\sc i}, as the suprathermal tail electrons will have little effect on
transitions for which the threshold energy is much lower than the
energy at which (locally) the power law tail of the EVDF begins. 
In such a case, even if the collision cross-section is larger at
energies in the power law tail, the rapid decline of the EVDF above
threshold (see Figure \ref{fig:4a}) means that excitation will be
dominated by the Maxwellian part of the distribution.
Hence the criterion for a transition to be relatively
unaffected by the presence of an enhanced suprathermal tail of the
type studied here is for its energy $E_{ij}$ to satisfy the inequality
\begin{equation}
E_{ij} < 0.1\xi_{\textrm{tail}}^{2} kT
\end{equation}
where $\xi_{\textrm{tail}}^{2} kT$ (= $mv_{\textrm{t}}^{2}/2$) is the
energy at which the suprathermal tail departs from the Maxwellian
distribution at temperature $T$. Even in the most extreme cases
encountered in the models tested here (i.e.\ $T \simeq 4000$ K,
$\xi_{\textrm{tail}} = 3.5$), this inequality is satisfied for
transitions in He~{\sc i} with $\Delta n = 0$ except for those with $n
= 2$, and for transitions between He~{\sc i} levels with $n = 4$ and
$n = 5$. Collisional excitation rates for transitions satisfying the
inequality are calculated assuming a Maxwellian distribution. 

\section{Results with non-Maxwellian collision rates}
\label{sec5}
\begin{figure}
\centering
\begin{minipage}{4in}
\epsfxsize=3.6in \epsfbox{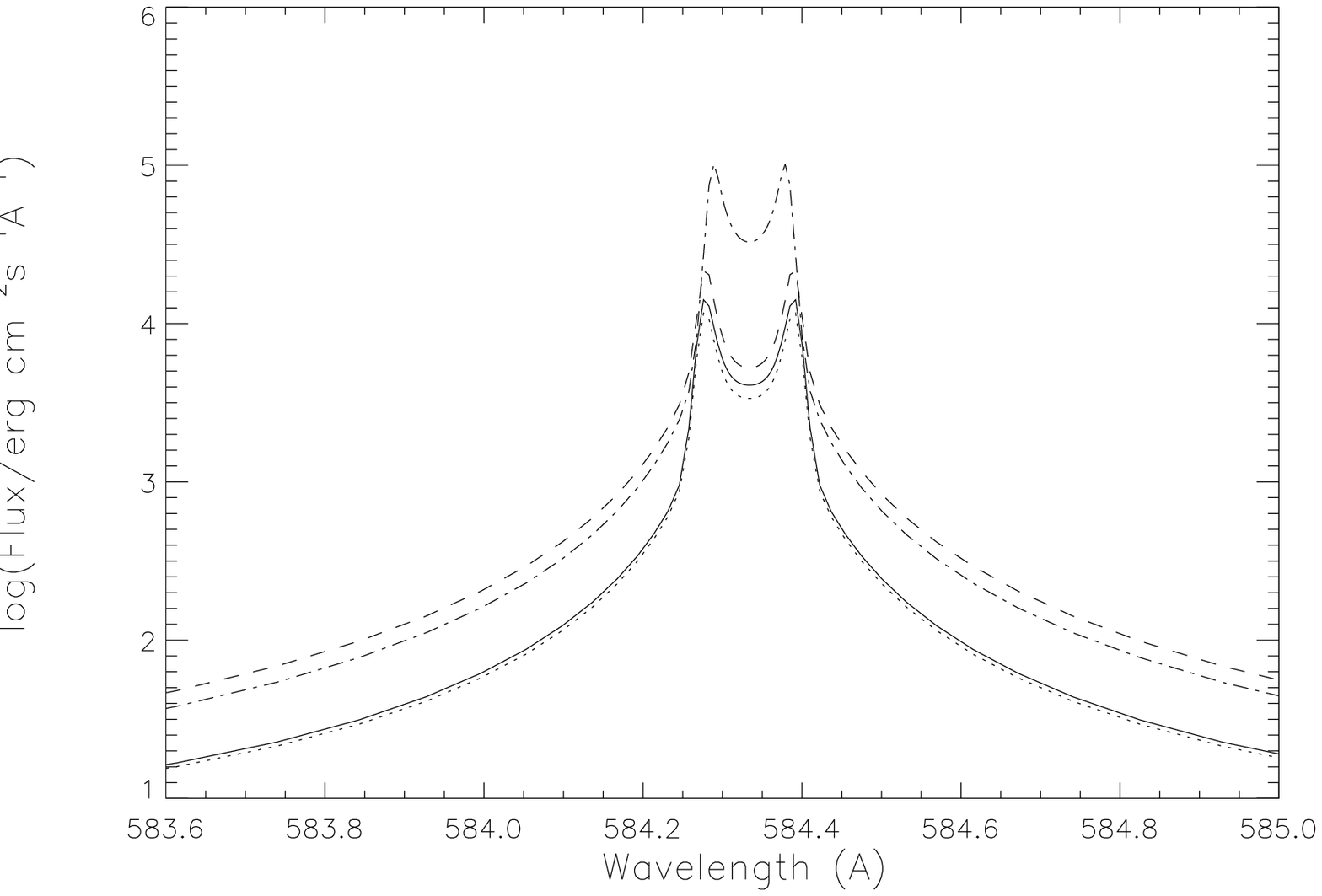}
\end{minipage}
\begin{minipage}{4in}
\epsfxsize=3.6in \epsfbox{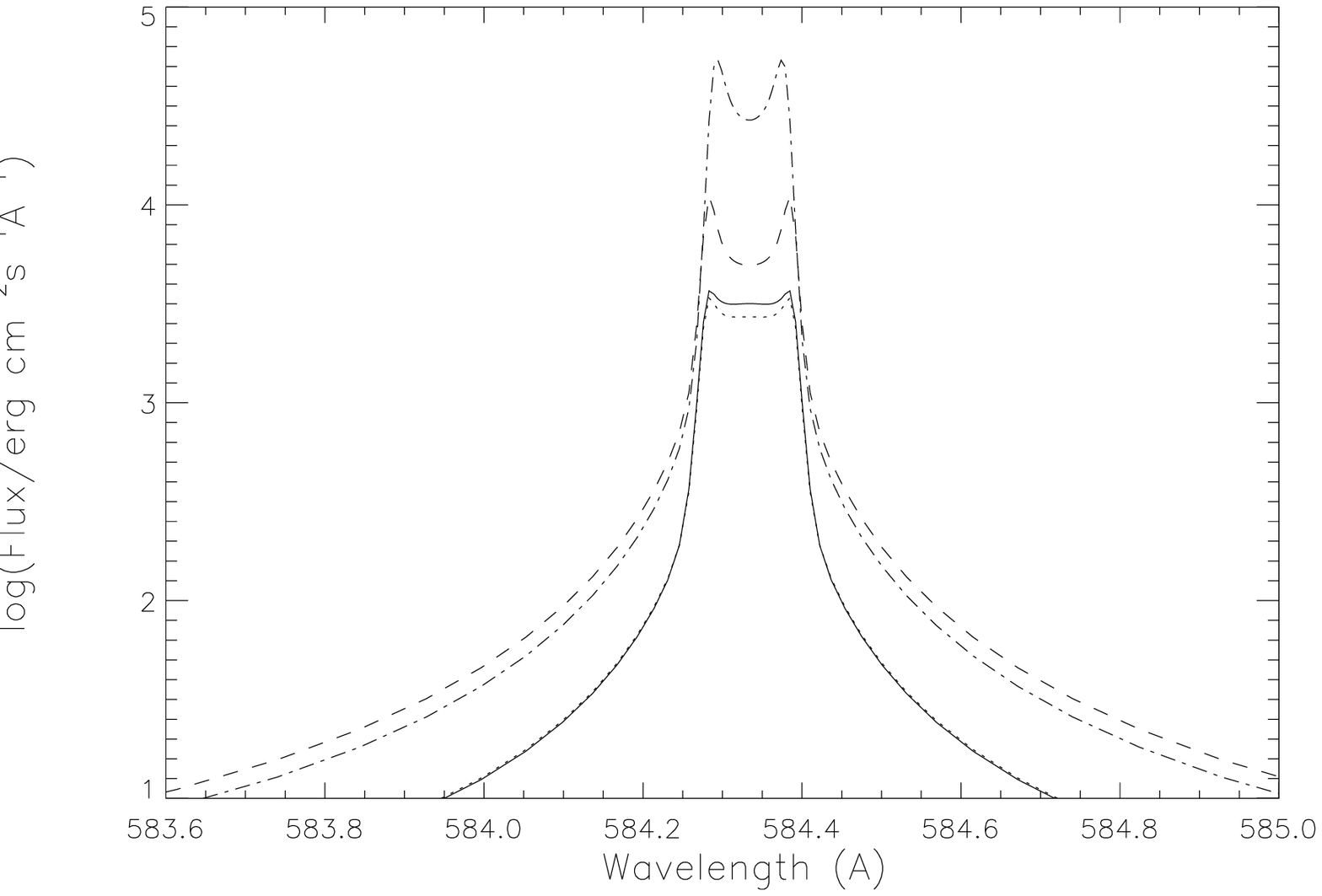}
\end{minipage}
\begin{minipage}{4in}
\epsfxsize=3.6in \epsfbox{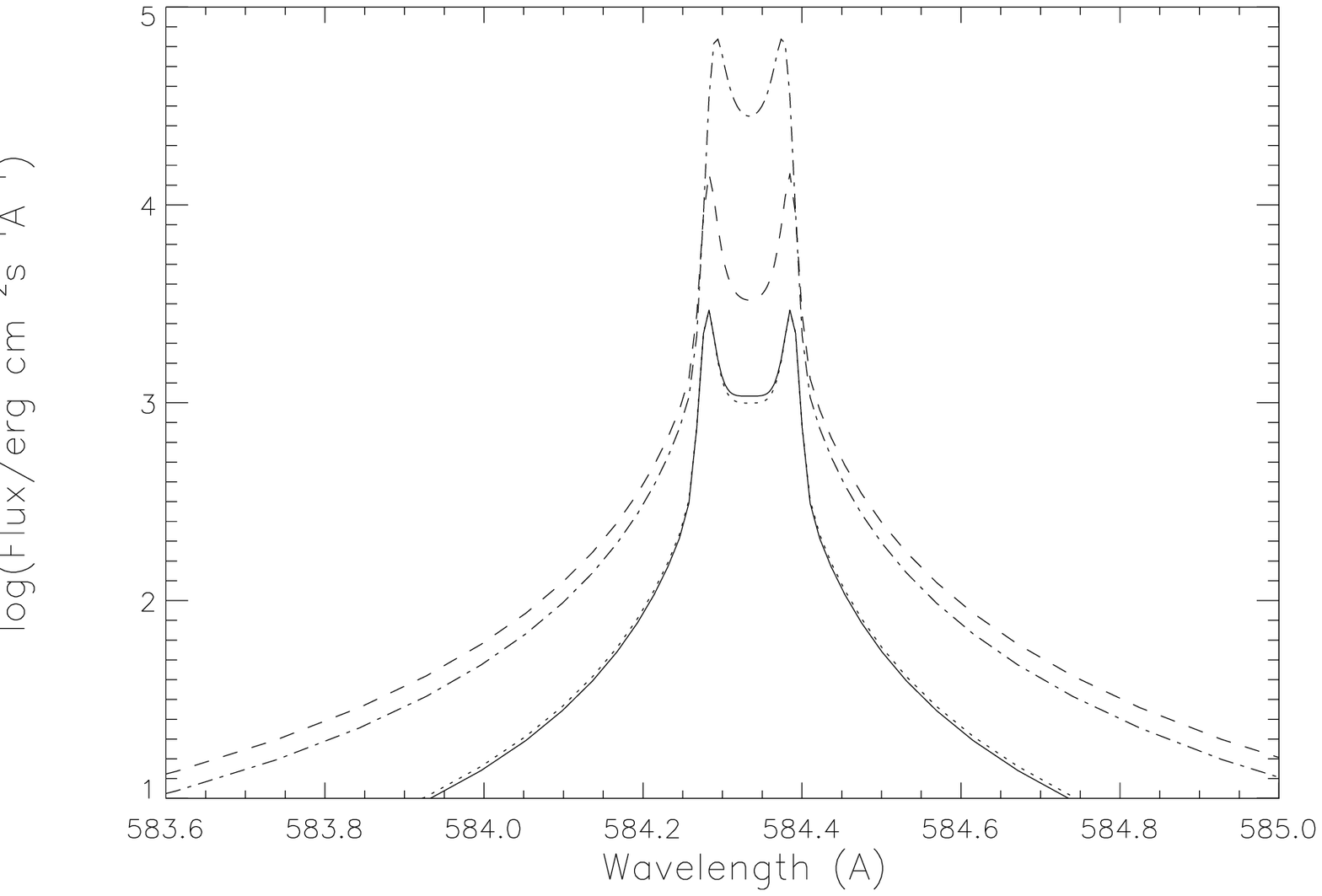}
\end{minipage}
\caption{He~{\sc i} 584.3-\AA\ line profiles produced by models VAL C
(top), and network S (middle) and X (bottom), for non-Maxwellian EVDFs with
$r_{\textrm{tail}} = 3.5$ and $\xi_{\textrm{tail}} = 5.0$ (solid),
$\xi_{\textrm{tail}} = 4.5$ (dotted), $\xi_{\textrm{tail}} = 4.0$
(dashed), $\xi_{\textrm{tail}} = 3.5$ (dot-dashed). \label{figX}}
\end{figure}

\begin{figure}
\centering
\begin{minipage}{3.6in}
\epsfxsize=3.6in \epsfbox{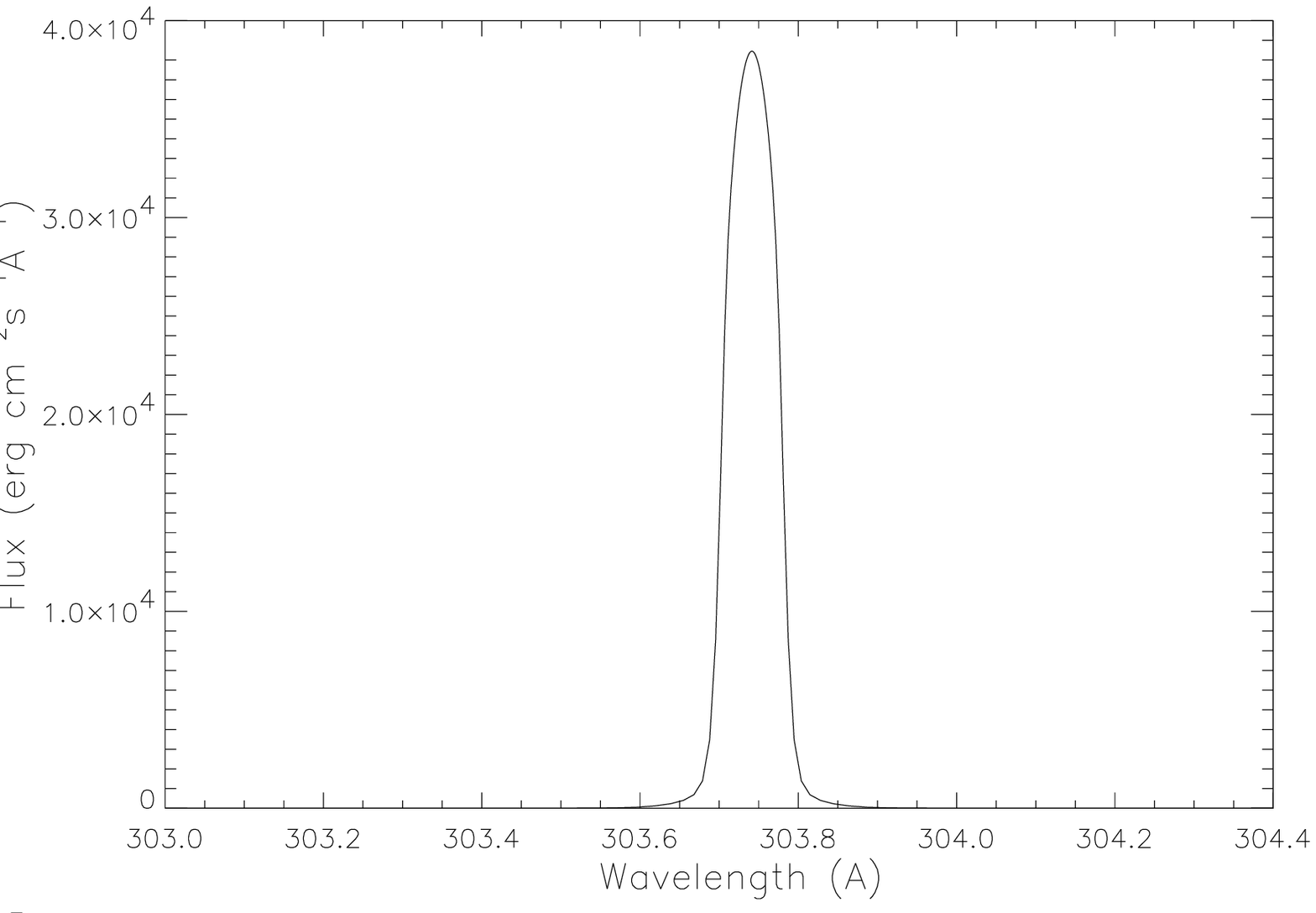}
\end{minipage}
\begin{minipage}{3.6in} 
\epsfxsize=3.6in \epsfbox{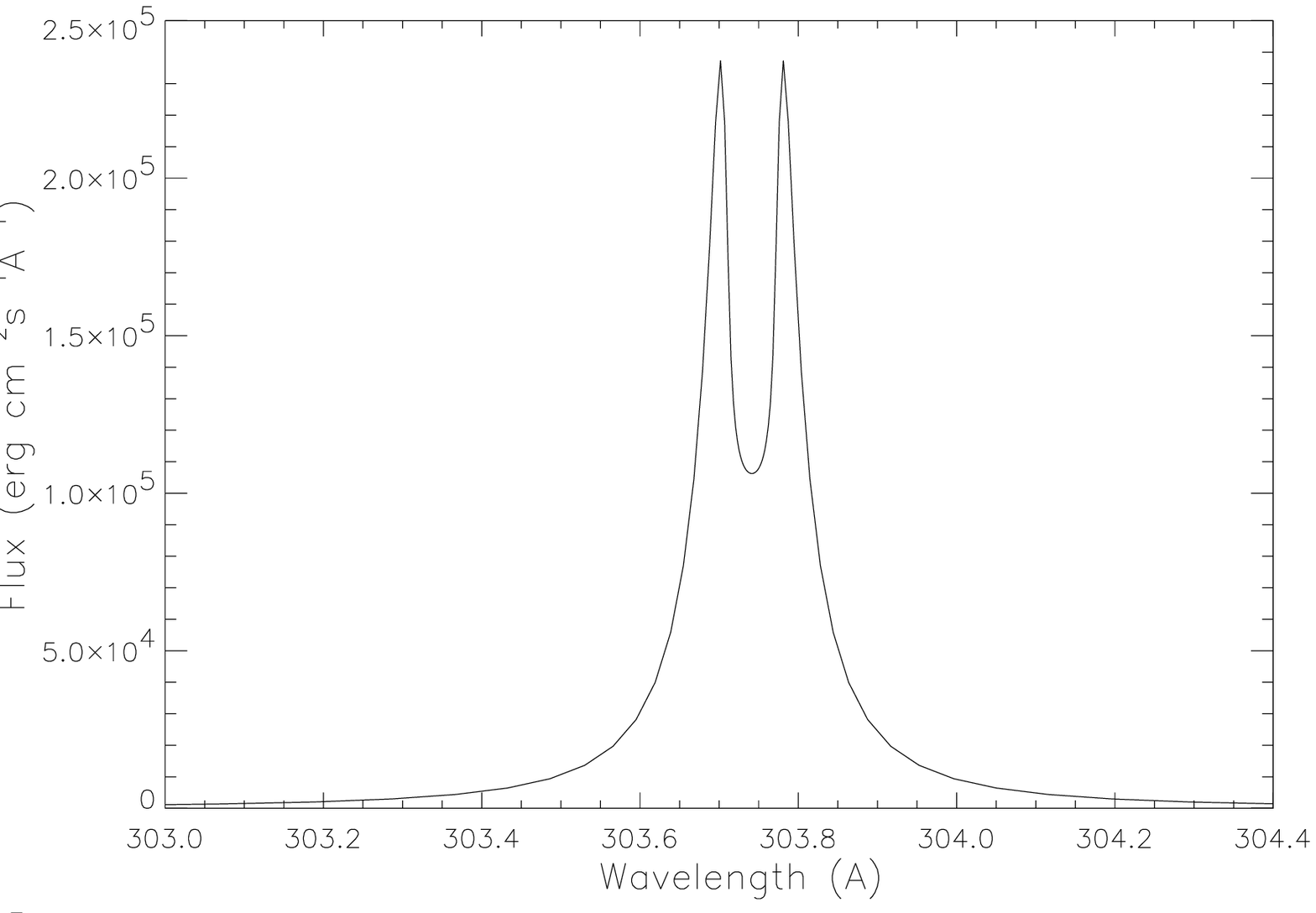}
\end{minipage}
\caption{He~{\sc ii} 303.8-\AA\ line profiles produced using network
model S, with Maxwellian collision rates (top), and with a
non-Maxwellian EVDF with $r_{\textrm{tail}} = 3.5, \xi_{\textrm{tail}}
= 3.5$ (bottom). \label{figY}} 
\end{figure}

The integrated intensities computed for the 303.8-\AA, 584.3-\AA, and
537.0-\AA\ lines are given in Tables \ref{tab3} -- \ref{tab6} 
for the VAL C model atmosphere and the network models S and X, 
for different values of the parameters $\xi_{\textrm{tail}}$ and
$r_{\textrm{tail}}$. Profiles of the 584.3-\AA\ and 303.8-\AA\ lines
computed for different values of $\xi_{\textrm{tail}}$, with
$r_{\textrm{tail}}$ = 3.5, are presented in Figures \ref{figX} and
\ref{figY}. 

Results for $\xi_{\textrm{tail}}$ = 3.5 are given only where
$r_{\textrm{tail}}$ = 3.5 
and 3.0, in order to test the EVDFs computed by Shoub
(1982,1983). This value of $\xi_{\textrm{tail}}$ is probably
unrealistic in the lower TR, given the calculations of Ljepojevic \&
Burgess \shortcite{lb90}, whose results are approximated here by
$r_{\textrm{tail}}$ = 2.0 and 4.5, $\xi_{\textrm{tail}} \geq 4.0$. The
results for $\xi_{\textrm{tail}}$ = 3.5 should be regarded with
caution, as the form of the 
modifications to the radiative transfer code results in there
being significant effects in chromospheric regions of the model
atmospheres, where in reality high energy electrons from the upper TR
are unlikely to penetrate. Non-Maxwellian tails might exist at these
heights if high energy electrons were accelerated \emph{locally} by MHD
processes, but without detailed models it is not known if the
resultant distributions would resemble the ones used here.

 \begin{table}
 \caption{Intensities and ratios of helium lines
 calculated using non-Maxwellian EVDFs with $r_{\textrm{tail}} =
 4.5$.\label{tab3}} 
 \begin{center}
 \begin{tabular}{ccccc}
 \hline
  & $\xi_{\textrm{tail}}$: & 5.0 & 4.5 & 4.0 \\\hline
 Model & $\lambda$ & \multicolumn{3}{c}{Computed intensity} \\
 atmosphere & (\AA) & \multicolumn{3}{c}{(erg cm$^{-2}$
 s$^{-1}$ sr$^{-1}$)} \\\hline
 VAL C & 584.3 & 400 & 329 & 675 \\
       & 537.0 & 51.9 & 99.7 & 192 \\
       & 303.8 & 2026 & 2032 & 3259 \\\hline
 Network S & 584.3 & 148 & 130 & 349 \\
	   & 537.0 & 26.7 & 48.4 & 134 \\
	   & 303.8 & 1165 & 1165 & 1776 \\\hline
 Network X & 584.3 & 84.3 & 85.3 & 392 \\
	   & 537.0 & 11.4 & 23.4 & 129 \\
	   & 303.8 & 994 & 999 & 1792 \\\hline
  & & \multicolumn{3}{c}{$I(\lambda)$/$I$(584.3 \AA)} \\\hline
 VAL C & 537.0 & 0.130 & 0.303 & 0.284 \\
       & 303.8 & 5.07 & 6.18 & 4.83 \\\hline
 Network S & 537.0 & 0.180 & 0.372 & 0.384 \\
	   & 303.8 & 7.87 & 8.96 & 5.09 \\\hline
 Network X & 537.0 & 0.135 & 0.274 & 0.329 \\
	   & 303.8 & 11.8 & 11.7 & 4.57 \\\hline
 \end{tabular}
 \end{center}
 \end{table}

 \begin{table}
 \caption{Intensities and ratios of helium lines
 calculated using non-Maxwellian EVDFs with $r_{\textrm{tail}} =
 3.5$.\label{tab4}}
 \begin{center}
 \begin{tabular}{cccccc}
 \hline
  & $\xi_{\textrm{tail}}$: & 5.0 & 4.5 & 4.0 & 3.5 \\\hline
 Model & $\lambda$ & \multicolumn{4}{c}{Computed intensity} \\
 atmosphere & (\AA) & \multicolumn{4}{c}{(erg cm$^{-2}$
 s$^{-1}$ sr$^{-1}$)} \\\hline
 VAL C & 584.3 & 419 & 349 & 681 & 1925\\
       & 537.0 & 47.2 & 95.2 & 181 & 699\\
       & 303.8 & 2026 & 2029 & 2272 & 29300\\\hline
 Network S & 584.3 & 150 & 135 & 323 & 1095 \\
	   & 537.0 & 24.6 & 45.8 & 118 & 479 \\
	   & 303.8 & 1165 & 1165 & 1262 & 13710 \\\hline
 Network X & 584.3 & 84.7 & 83.6 & 336 & 1320 \\
	   & 537.0 & 10.3 & 21.0 & 102 & 536 \\
	   & 303.8 & 994 & 994 & 1107 & 17310 \\\hline
  & & \multicolumn{4}{c}{$I(\lambda)$/$I$(584.3 \AA)} \\\hline
 VAL C & 537.0 & 0.113 & 0.273 & 0.266 & 0.363 \\
       & 303.8 & 4.84 & 5.81 & 3.34 & 15.2 \\\hline
 Network S & 537.0 & 0.164 & 0.339 & 0.365 & 0.437 \\
	   & 303.8 & 7.77 & 8.63 & 3.91 & 12.5 \\\hline
 Network X & 537.0 & 0.122 & 0.251 & 0.304 & 0.406 \\
	   & 303.8 & 11.7 & 11.9 & 3.29 & 13.1 \\\hline
 \end{tabular}
 \end{center}
 \end{table}

 \begin{table}
 \caption{Intensities and ratios of helium lines
 calculated using non-Maxwellian EVDFs with $r_{\textrm{tail}} =
 3.0$.\label{tab5}} 
 \begin{center}
 \begin{tabular}{cccccc}
 \hline
  & $\xi_{\textrm{tail}}$: & 5.0 & 4.5 & 4.0 & 3.5 \\\hline
 Model & $\lambda$ & \multicolumn{4}{c}{Computed intensity} \\
 atmosphere & (\AA) & \multicolumn{4}{c}{(erg cm$^{-2}$
 s$^{-1}$ sr$^{-1}$)} \\\hline
 VAL C & 584.3 & 411 & 357 & 639 & 2243\\
       & 537.0 & 44.7 & 90.6 & 179 & 800\\
       & 303.8 & 2025 & 2028 & 2145 & 12200\\\hline
 Network S & 584.3 & 152 & 139 & 308 & 1194 \\
	   & 537.0 & 23.7 & 43.8 & 113 & 561 \\
	   & 303.8 & 1165 & 1165 & 1227 & 5709 \\\hline
 Network X & 584.3 & 86.1 & 84.7 & 313 & 1527 \\
	   & 537.0 & 9.87 & 19.9 & 93.7 & 606 \\
	   & 303.8 & 994 & 994 & 1048 & 6691 \\\hline
  & & \multicolumn{4}{c}{$I(\lambda)$/$I$(584.3 \AA)} \\\hline
 VAL C & 537.0 & 0.109 & 0.254 & 0.280 & 0.357 \\
       & 303.8 & 4.93 & 5.68 & 3.36 & 5.44 \\\hline
 Network S & 537.0 & 0.156 & 0.315 & 0.367 & 0.470 \\
	   & 303.8 & 7.66 & 8.38 & 3.98 & 4.78 \\\hline
 Network X & 537.0 & 0.115 & 0.235 & 0.299 & 0.397 \\
	   & 303.8 & 11.5 & 11.7 & 3.35 & 4.38 \\\hline
 \end{tabular}
 \end{center}
 \end{table}

 \begin{table}
 \caption{Intensities and ratios of helium lines
 calculated using non-Maxwellian EVDFs with $r_{\textrm{tail}} =
 2.0$.\label{tab6}}
 \begin{center}
 \begin{tabular}{ccccc}
 \hline
  & $\xi_{\textrm{tail}}$: & 5.0 & 4.5 & 4.0 \\\hline
 Model & $\lambda$ & \multicolumn{3}{c}{Computed intensity} \\
 atmosphere & (\AA) & \multicolumn{3}{c}{(erg cm$^{-2}$
 s$^{-1}$ sr$^{-1}$)} \\\hline
 VAL C & 584.3 & 410 & 380 & 576 \\
       & 537.0 & 40.7 & 75.9 & 226 \\
       & 303.8 & 2025 & 2025 & 2049 \\\hline
 Network S & 584.3 & 153 & 147 & 248 \\
	   & 537.0 & 22.0 & 36.8 & 115 \\
	   & 303.8 & 1165 & 1165 & 1179 \\\hline
 Network X & 584.3 & 89.0 & 89.3 & 233 \\
	   & 537.0 & 9.09 & 16.7 & 87.2 \\
	   & 303.8 & 994 & 994 & 998 \\\hline
  & & \multicolumn{3}{c}{$I(\lambda)$/$I$(584.3 \AA)} \\\hline
 VAL C & 537.0 & 0.099 & 0.200 & 0.392 \\
       & 303.8 & 4.94 & 5.33 & 3.56 \\\hline
 Network S & 537.0 & 0.144 & 0.250 & 0.463 \\
	   & 303.8 & 7.61 & 7.93 & 4.75 \\\hline
 Network X & 537.0 & 0.102 & 0.187 & 0.374 \\
	   & 303.8 & 11.2 & 11.1 & 4.28 \\\hline
 \end{tabular}
 \end{center}
 \end{table}

Even if the results for $\xi_{\textrm{tail}}$ = 3.5 are discounted,  
comparison of Tables \ref{tab3} -- \ref{tab6} with Table \ref{tab2}
shows that collisional excitation by non-Maxwellian electron
distributions can lead to significant enhancements of the helium resonance
line intensities over those when Maxwellian distributions are
assumed. The results also show that non-Maxwellian EVDFs could be
present in the TR without producing a strong signal in the helium
lines (compare Tables \ref{tab6} and \ref{tab2}). Calculations for 
$\xi_{\textrm{tail}} = 5.0, r_{\textrm{tail}} = 2.0$, parameters
characteristic of  
EVDFs computed by Ljepojevic \& Burgess \shortcite{lb90}, suggest that
the only observable effect of such a distribution would be an increase in
the intensity of the He~{\sc i} 537.0-\AA\ line with respect to the
case of Maxwellian excitation.

The trends in the calculated intensities are largely
understandable. For a given $r_{\textrm{tail}}$, a reduction of
$\xi_{\textrm{tail}}$ generally causes increases in line
intensities, as the power law tail becomes significant at lower
electron energies, and at a given temperature increases the proportion
of electrons with energies high enough to collisionally excite a given
line. A steeper power law (smaller $r_{\textrm{tail}}$) tends to
suppress He~{\sc ii} with respect to He~{\sc i} (compare Tables
\ref{tab3} and \ref{tab6}), as higher energy electrons are present in
relatively smaller numbers. 

For any particular $r_{\textrm{tail}}$, the He~{\sc ii} resonance line
intensity shows a sharper increase at low $\xi_{\textrm{tail}}$ than
do the He~{\sc i} lines, 
which is due to a significant change in the ionization
balance. Enhanced line emission can occur, not just because of the
increased collisional excitation of the relevant transition, but also
because the shape of the peak in the ionization fraction can be
changed by the enhanced ionization rates at low temperatures. Thus
emission occurs over a larger, more dense region in He~{\sc ii}, whereas
the region of peak He~{\sc i} ionization fraction can only be narrowed. At
the lower limit of the $\xi_{\textrm{tail}}$ tested, the ionization fractions
at a given temperature depart from the standard model results by factors
approaching, and in some places exceeding, an order of magnitude. As
stated above, this may be unrealistic in the regions of He~{\sc i}
line formation.

The formation of the He~{\sc ii} resonance line in higher density
regions when $\xi_{\textrm{tail}}$ is small can lead to its profile
becoming self-reversed. For $\xi_{\textrm{tail}} \geq 4.0$, 
the profile is approximately Gaussian, and very similar to that
predicted by calculations made with the non-Maxwellian
modifications. The comparsion is shown for model S in Figure
\ref{figY}; results from the other models are similar. Figure
\ref{figX} shows that the He~{\sc i} resonance line is self-reversed
in all cases of non-Maxwellian excitation.

The somewhat surprising \emph{reduction} of the 584.3-\AA\ line intensity
(compared to calculations with `normal' Maxwellian rates) for higher
$\xi_{\textrm{tail}}$ (compare Tables \ref{tab4} and \ref{tab2})
appears to be due to a similar effect. The shift in the ionization balance
suppresses the Maxwellian contribution to collisional excitation. The
effect is smallest in model X, where Maxwellian collisional excitation
is least significant because of the smaller amount of material in the
relevant temperature range. The 584.3-\AA\ line intensity can
be reduced if the tail has a value of $\xi_{\textrm{tail}}$ too high
for increased excitation by suprathermal electrons to
compensate. Other lines of the He~{\sc i} resonance series do not show
this, as the increased excitation has a greater effect than the shift
in ionization (the suppressed Maxwellian contribution is relatively
less significant). 

The effect of enhanced ionization on the ionization balance also
complicates the trend in the ratio $I$(537.0~\AA)/$I$(584.3~\AA),
which is expected to decrease with decreasing $r_{\textrm{tail}}$. For
low values of $\xi_{\textrm{tail}}$, the trend can reverse, as the
change in the ionization balance affects the 584.3-\AA\ line more than
the 537.0-\AA\ line, allowing $I$(537.0~\AA) to grow more quickly with
decreasing $\xi_{\textrm{tail}}$. 

\subsection{Comparison with observations}
The results of the calculations may be compared with
the mean line intensities observed by MJ99, which are given
in Table \ref{tab1}. It can be seen that no particular combination of
$\xi_{\textrm{tail}}$ and $r_{\textrm{tail}}$ leads to a good match to
observations in all three lines for any of the model atmospheres tested. 

In models in which the
calculated 303.8-\AA\ line intensity is comparable to observed
values, the He~{\sc i} line ratio $I$(537.0~\AA)/$I$(584.3~\AA) is
much larger than observed (independent of the {\sc cds} calibration
assumed), and the computed 303.8-\AA\ line profile is in general
self-reversed. Such reversals are not observed, but would probably not be
resolved by the instruments on board {\em SOHO} \cite{grs00}.
Models showing enhancements of the 303.8-\AA\ line intensity comparing
well with observations 
have other problems. The values of $\xi_{\textrm{tail}}$ required may
be unrealistically low, but if such such a tail \emph{were} present,
significant collisional enhancement of the He~{\sc ii} 1640.4-\AA\
multiplet is predicted. Intensities much larger than those
observed, in some cases by orders of magnitude, are produced when
$\xi_{\textrm{tail}} < 4.5$ (independent of $r_{\textrm{tail}}$). This
is in part due 
to the increased fraction of He~{\sc ii} at lower temperatures
where recombination dominates the formation of the multiplet. 
Small departures from Maxwellian EVDFs might explain the Balmer
$\alpha$ intensity, but results found here imply that
non-Maxwellian collisional excitation could not be responsible for the
observed resonance line intensity without producing too much Balmer
$\alpha$ emission. It remains to be seen whether non-Maxwellian collisional
excitation is consistent with observations of the profiles and relative
intensities of the components of the 1640.4-\AA\ multiplet. An
improved treatment of the energy levels of He~{\sc ii} would allow an
investigation of this point. 
The excitation of the 1640.4-\AA\ multiplet by non-local electrons
will also be tested against stellar observations in future work. 

Trends in the results given in Tables \ref{tab3} -- \ref{tab6}
may be compared with observed variations in line
intensities and ratios over different regions of the Sun. The
effectiveness of the excitation
mechanism depends on the temperature gradient and electron density,
which have different values in the network and internetwork, in the
quiet Sun and coronal holes. 
Given the significant disagreements between the calculations and
observations described above, an approximate approach was followed in
preference to constructing models of different parts of the atmosphere.
The effects of different atmospheric parameters
on the suprathermal electron tail are considered, and the results
given in Tables \ref{tab3} -- \ref{tab6} are used to infer how the
helium line intensities might vary as a result. 

Changes in the suprathermal tail may be parametrized in simple terms
as variations of $\xi_{\textrm{tail}}$ and $r_{\textrm{tail}}$. The
variation of $r_{\textrm{tail}}$ appears to depend on the energy
balance in the atmosphere (Shoub 1982), and given the current lack of
detailed understanding of the heating of the atmosphere, it is
difficult to predict how conditions in different regions
might affect $r_{\textrm{tail}}$. On the other hand, Shoub's
(1982) BGK calculations provide an expression for the variation of
$\xi_{\textrm{tail}}$ with easily understood physical quantities:  
\begin{equation}
\label{eq:4c}
\xi_{\textrm{tail}} \propto \left(\frac{T_{\textrm{h}}}{L}\right)^{-1/6}
\left(\frac{N_{\textrm{e}}(T_{\textrm{h}})}{T_{\textrm{h}}}\right)^{1/6}
\left(\frac{T}{T_{\textrm{h}}}\right)^{1/12}
\end{equation}
where $L$ is the thickness of the transition region and
$T_{\textrm{h}}$ is the temperature at its upper edge;
$T_{\textrm{h}}/L$ corresponds to the average temperature 
gradient. $\xi_{\textrm{tail}}$ is therefore decreased (and the
influence of the suprathermal tail increased) by an increase in
$T_{\textrm{h}}$ or the temperature gradient d$T/$d$h$, or by a
decrease in the electron pressure. This relation emerges from
analytical calculations in a simplified system, but it shows expected 
dependences. The mean 
free path of an electron with a given velocity should increase with
decreasing electron density, and the change in temperature over that
path will increase as the temperature gradient steepens. Thus such
changes increase the numbers of electrons from a region at $T_{1}$
reaching a region of lower temperature $T_{2}$. Increasing $T_{1}$
shifts the peak of the distribution function to higher energies,
increasing the number of electrons in the tail with a given
velocity. This also leads to an increase in the number of electrons
with mean free paths large enough to influence lower temperature
regions. 

The dependence of $\xi_{\textrm{tail}}$ on each of the parameters is
weak, but Tables \ref{tab3} -- \ref{tab6} show that the helium line
intensities can be sensitive to relatively small changes in
$\xi_{\textrm{tail}}$. If changes in $\xi_{\textrm{tail}}$ are assumed
to control the variation of helium line emission with position on the
Sun, the predicted variations in intensities may be tested against
observations. 

With respect to average conditions in the quiet Sun, in coronal holes
the electron pressure is 2 -- 3 times smaller, the mean temperature
gradient up to an order of magnitude smaller, and the coronal
temperature at least 30 per cent smaller \cite{mw72,jea01}. According to
the expression in equation (\ref{eq:4c}), $\xi_{\textrm{tail}}$ would
then be a factor of up to $\sim 1.5$ larger in coronal holes than in
the quiet Sun. Such a change spans the values investigated here; Tables
\ref{tab3} -- \ref{tab6} show that over such a range, very large
changes in the helium line intensities can occur. 
The helium resonance line intensities are \emph{observed} to be
factors of 1.5 -- 2.0 smaller in coronal holes than in the quiet Sun
\cite{hp99,jea01}. The present calculations suggest that such a
decrease in He~{\sc i} line intensities would be accompanied by a
large decrease in the $I$(537.0~\AA)/$I$(584.3~\AA) line ratio
(perhaps by a factor of two or more). This would outweigh the small
increase in the ratio expected with the reduction of coronal
illumination. Observations \cite{jea01} suggest that the ratio is
slightly larger in coronal holes than in the quiet Sun, contrary to
the predictions of models dominated by non-Maxwellian collisional
excitation. Changes in $\xi_{\textrm{tail}}$ of a 
factor of 1.5 also generally lead to significant variations in the
calculated ratio of the He~{\sc i} and He~{\sc ii} resonance lines. In
contrast, observations show no consistent variation in this ratio
between coronal holes and the quiet Sun; the mean observed ratio is
almost exactly the same in the two regions \cite{jea01}. 

The observed trends in line ratios in the quiet Sun may also be
compared with the predictions of the calculations. In the quiet Sun
MJ99 observed only small changes in the ratio
$I$(537.0~\AA)/$I$(584.3~\AA) for changes in $I$(584.3~\AA) of a
factor of 5, the ratio decreasing with increasing $I$(584.3~\AA). The
ratio $I$(303.8~\AA)/$I$(584.3~\AA) is also observed to vary by
relatively small amounts, and while consistent trends are not seen
clearly, small values of the ratio often coincide with large values of
$I$(584.3~\AA). For large variations in $I$(584.3~\AA), the
calculations predict $I$(537.0~\AA)/$I$(584.3~\AA) to \emph{increase}
with increasing $I$(584.3~\AA) in all but a small part of the
parameter space, particularly for the (probably more realistic)
smaller values of $r_{\textrm{tail}}$. The variation of the ratio is also
larger than in observations. The predicted trend appears inevitable
given the exciting mechanism, since the power law tail provides
relatively greater numbers of electrons (with respect to a Maxwellian
distribution) at the excitation energy of the 537.0-\AA\ line
than at the energy of the 584.3-\AA\ line. The models also predict
much larger variations in the ratio $I$(303.8~\AA)/$I$(584.3~\AA) than
are observed for given changes in $I$(584.3~\AA). Thus the predicted
variations of the helium line ratios are \emph{not} consistent
with observations.

The comparisons made so far with quiet Sun observations have ignored
the question of whether $\xi_{\textrm{tail}}$ would be expected to
vary in a manner which would produce the observed variations in
intensities between cell interior and network regions.
In considering the dependence of $\xi_{\textrm{tail}}$ on the
conditions in the quiet Sun transition region, the likely magnetic
field geometry should be taken into account together with the changes
in temperature  gradient and electron pressure. All calculations
described above assume any magnetic field to be aligned with a radial
temperature gradient. In general, electrons streaming from the corona
and upper TR would tend to follow the magnetic field, which in the
lower TR and chromosphere is concentrated in the network. The model of
Gabriel \shortcite{ag76} suggests that the temperature gradient
\emph{in the direction of the magnetic field} is largest in the centre
of the network and decreases towards the edges. The model also predicted
lower pressures in the network, while MJ99 observed such a
trend. Both factors would cause $\xi_{\textrm{tail}}$ to be smallest at the
centres of network boundaries, suggesting that enhancement of the
helium line intensities by non-local electrons would be concentrated
in the centre of the network. Such
behaviour is not observed; indeed the helium network appears to be
unusually \emph{broad} (e.g.\ Brueckner \& Bartoe 1974; Gallagher et
al.\ 1998), and the enhancement of the
helium lines with respect to other transition region lines appears to
increase towards cell interiors (MJ99). This observed behaviour could be
due to radiative transfer effects not considered in this simple
argument, but it presents further problems for the proposed
enhancement process. In contrast, it is argued in Paper I that
enhancement by upward transport of helium ions in an expanding network
field would lead naturally to the network appearing wider in the
helium resonance lines than in other low TR lines.

\subsection{Approximations in the EVDF}
Given that the main approximation in the present work is that the form
of the EVDF is assumed rather than calculated self-consistently, the
extent to which the form chosen for the EVDF affects the computed line
intensities should be considered. Two features of the EVDF in
particular were investigated: the behaviour of the tail at very high
velocities, and the dependence (or otherwise) of $\xi_{\textrm{tail}}$ on
temperature. 

Shoub's \shortcite{es82} analytical results show a down-turn in the high
energy tail, where the power law ceases to be a good description, at
$\xi \sim 4.2 \times 10^{3} T^{-1/2}$. The numerical calculations do
not show this feature, these values of $\xi$ being
outside the range of calculation, but it is plausible that such a
down-turn should exist, as the above limit on $\xi$ decreases with
increasing $T$, so that the EVDF tends towards the Maxwellian
distribution assumed to exist in the near-isothermal corona. 
When calculations were made with a cut-off in the EVDF at the above
$\xi(T)$, significant effects on the helium resonance lines were found
for small $\xi_{\textrm{tail}}$. The decreased extent of the power law
tail at higher temperatures reduces ionization rates from He~{\sc i} to
He~{\sc ii}, decreasing the extent of the region where He~{\sc ii}
dominates the ionization balance. Consequently, the intensity of the
303.8-\AA\ line is decreased by up to a factor of 2, and the
intensities of the He~{\sc i} lines are increased by up to 50 per cent. The
cut-off is a crude device and calculations with $r_{\textrm{tail}} < 3.5$ may
well approximate the effects of the down-turn more accurately, but the
sensitivity revealed here suggests that this part of the distribution
may need closer attention in the future. The present results imply
that including the down-turn properly in calculations would not
improve the fit to observations. 

The use in this study of EVDFs with $\xi_{\textrm{tail}}$ independent
of $T$ is an approximation suggested by the general form of solutions
to the Boltzmann equation. Shoub's \shortcite{es82} analytical
solutions suggest $\xi_{\textrm{tail}} \sim 
T^{1/12}$ (thus it changes by only 20 per cent for an order of
magnitude change in $T$). Including such a dependence in calculations
would be expected to worsen agreement with observations. The observed
He~{\sc ii} resonance line intensity can only be produced by low
values of $\xi_{\textrm{tail}}$, but the ratio of the 584.3-\AA\ and
537.0-\AA\ lines favours higher values, which would require
$\xi_{\textrm{tail}}$ to vary with $T$ in the opposite
sense. Ljepojevic \& Burgess' \shortcite{lb90} results \emph{do} show
some increase in $\xi_{\textrm{tail}}$ for the lowest temperatures for
which they present the EVDF. If $\xi_{\textrm{tail}}$ were lower at
the higher $T$ of He~{\sc ii} line formation than at the lower $T$ of
He~{\sc i} line formation, a better fit might be obtained, although
the problem with the He~{\sc ii} Balmer $\alpha$ intensity would
remain. This problem is certainly seen in the results of calculations
made with enhanced collision rates operating only in He~{\sc ii} and
not in He~{\sc i}, approximating very roughly the situation in which
$\xi_{\textrm{tail}}$ is increased at low temperatures.

\section{Summary and conclusions}
\label{sec6}
A detailed study has been made of the formation of the helium resonance
lines, involving radiative transfer calculations with a 36 level
model helium atom. Calculations using the the VAL C (average quiet
Sun) model atmosphere, with a coronal radiation field characteristic
of the quiet Sun based on the solar irradiance model of Tobiska
\shortcite{kt91}, come close to reproducing the observed He~{\sc i}
584.3-\AA\ line intensity, but produce intensities in the He~{\sc i}
537.0-\AA\ and He~{\sc ii} 303.8-\AA\ lines smaller than are
observed. The predicted self-reversal of the 584.3-\AA\ line is not
observed, but it is shown that it may be largely obscured by instrumental
effects. Calculations using the VAL D (network) model produce
results in better agreement with observed network intensities, but the
VAL models both produce poor matches to other transition region lines
(e.g.\ of C~{\sc ii} and Si~{\sc iii}).

New model atmospheres (S and X) appropriate to network boundary
regions were constructed to be more consistent with observed
intensities of TR lines of species other than helium. 
Calculations using these new models with quiet coronal illumination
fail to reproduce the observed quiet Sun network He~{\sc ii} resonance
line intensity by factors of 7--8 (3--4 when the most recent {\sc
cds} calibration is used). Model X, based on the $\xi$ Boo A
EMD, produces an intensity for the He~{\sc i} resonance line a factor
of 7.5 smaller than observed, but model S, based on the MJ99 solar network
EMD, gives better agreement. The computed intensity for the 584.3 \AA\
line is a factor of 4 too small, but the computed line profile is
closest to the observed one, having only a small self-reversal that
could be completely obscured by instrumental broadening. The ratio 
$I$(537.0~\AA)/$I$(584.3~\AA) computed using model S also matches
observations most closely.

The optical thickness of the He~{\sc i} and He~{\sc ii} resonance
lines may explain the unusual width of the network observed in those
lines if photons are scattered from the edges of the network. Studies
of two-component model atmospheres are required to investigate this point.

In all of the model atmospheres the formation of the He~{\sc ii}
resonance line is dominated by collisional excitation, but
recombination is more important in the formation of the He~{\sc ii}
1640.4-\AA\ multiplet. Recombination also contributes significantly to
the intensity of the He~{\sc i} 584.3-\AA\ line, but does not dominate
its formation except in model X. Direct collisional excitation from
the ground and indirect excitation by radiative transitions from 
other collisionally excited singlet levels also contribute
significantly to the formation of the line. Similar formation
processes are important in the He~{\sc i} 537.0-\AA\ line.

Increasing the intensity of the coronal radiation field causes
increases in the computed He~{\sc i} line intensities and generally produces
a decrease in the $I$(537.0~\AA)/$I$(584.3~\AA) ratio. The He~{\sc ii}
resonance line intensity is generally \emph{reduced} by increased coronal
illumination, as a shift in the peak He~{\sc ii} ionization fraction to
lower temperatures reduces collisional excitation rates. Removing the
coronal illumination produces very little change in the computed
He~{\sc ii} resonance line intensity. An
absence of coronal radiation does cause a reduction in the computed
intensities of the He~{\sc i} lines, an effect approaching the observed
reduction in model S, exceeding it in model X. The line ratio
$I$(537.0~\AA)/$I$(584.3~\AA) is observed to be slightly larger in coronal
holes than in the quiet Sun, and this feature is reproduced by the
models with zero coronal illumination. The quite different responses
to changes in coronal 
irradiance of the He~{\sc i} and He~{\sc ii} resonance lines suggest
that this is not the factor controlling factor the intensities of both
lines, given that they show similar behaviour, both in
the quiet Sun and coronal holes, and that the ratio of their observed
intensities does not change by large amounts (MJ99, Jordan et al.\
2001). 

Given that the intensities of the helium lines with respect to other
TR lines are not explained by the `standard' models used in the
present work, modifications were made to the radiative transfer code
to allow the effects of non-Maxwellian EVDFs on
collision rates to be included in calculations of the helium line
intensities. The EVDFs tested are of a form postulated to exist in the
transition region because of the extreme temperature gradient (e.g.\
Shoub 1983); they are Maxwellian at thermal velocities, with a power
law tail at suprathermal velocities. The parameters of the power law
tail were varied in a range suggested by the results of Shoub
(1982,1983) and Ljepojevic \& Burgess \shortcite{lb90}, allowing an
exploration of EVDFs that might plausibly be present in the model
atmospheres studied here.

It is found that this simulation of the presence of an increased
population of suprathermal electrons in the lower transition region
can lead to significant enhancements of the helium resonance line
intensities compared with models assuming Maxwellian electron
distributions. None of the models tested, however, simultaneously
reproduce the observed intensities of the three lines
studied. Calculations in which the He~{\sc ii} resonance line
intensity is approximately reproduced also produce a much
larger intensity in the He~{\sc ii} 1640.4-\AA\ multiplet than is
observed. The enhanced collision rates produce large values of
the He~{\sc i} line ratio $I$(537.0~\AA)/$I$(584.3~\AA), as the
increase over Maxwellian rates is relatively greater in the higher
energy transition. In some cases, where the power law tail is relatively
insignificant, this produces better matches between the computed and
observed ratios, but in cases where the suprathermal tail dominates
excitation the computed ratio is generally much larger than
observed. The calculations also predict that the ratio would increase
with the absolute intensities of the lines, which is the opposite of
the observed trend (MJ99).

A consideration of how local atmospheric conditions might affect the
EVDF at a given temperature in the low TR suggests that a reduction of
the helium resonance line intensities in coronal holes with respect to
the quiet Sun \emph{would} be expected if excitation by non-local
electrons dominates line formation. However, the models also predict
that a decrease in absolute He~{\sc i} intensities of the observed
order would be accompanied by a decrease in the line ratio
$I$(537.0~\AA)/$I$(584.3~\AA); the opposite is observed \cite{jea01}. 
This excitation process would also probably produce a larger variation
in the ratio $I$(303.8~\AA)/$I$(584.3~\AA) than is observed between
coronal holes and the quiet Sun \cite{jea01}. Similar considerations
suggest that the appearance of the network in the helium lines is not
easily explained in the case of excitation by non-local electrons. 

In general, the calculations predict that significant departures from
Maxwellian EVDFs of the form investigated would produce signatures in
the helium line ratios which contradict observations. Small departures
from Maxwellian EVDFs are not ruled out, but I conclude that
collisional excitation by suprathermal electrons in EVDFs of the form
examined here is unlikely to dominate the formation of the helium
resonance lines. Enhancement by non-thermal transport of helium atoms
and ions, as investigated in Paper I, appears to be a more promising
explanation of the helium resonance line intensities.

The calculations performed in the present work contain a number of
approximations, and the above conclusions are therefore contingent on
a more sophisticated study of the excitation process.
Ideally, such an investigation would be totally self-consistent,
solving the Boltzmann equation for the EVDF and the equation of
radiative transfer simultaneously. Models in which suprathermal electron
populations are generated by reconnection events or MHD processes in
the low TR (rather than by the steep TR temperature gradient) also
deserve examination. The apparent sensitivity of the He~{\sc ii}
1640.4 \AA\ multiplet to the presence of suprathermal electrons (in
part through the altered ionization balance) suggests that these lines
also deserve further investigation. In particular, the plausibility of
enhanced suprathermal excitation could be tested against the observed
profile of the multiplet and also against stellar observations.

\section*{Acknowledgments}
I would like to thank Prof.\ Carole Jordan for her advice during this
work. I also gratefully acknowledge the financial support of PPARC as
a DPhil student, under grant PPA/S/S/1997/02515.

\section*{Appendix}
The new model atmospheres described in Section \ref{sec2.2} are
presented in the following pages. The tables give the logarithm (to base 10)
of the mass column density $m$ (in units of g cm$^{-2}$), the electron
temperature, the neutral hydrogen, proton, and electron number
densities, the microturbulent velocity $v_{\textrm{\sc t}}$, and the
height, $h$, above optical depth unity at 5000 \AA.

\newpage

\begin{table}
\caption{Network model S}\label{tabA7}
\begin{center}
\begin{tabular}{ccccccc}
\hline
 log $m$ & $T_{e}$ (K) & $N_{H}$ (cm$^{-3}$) & $N_{p}$ (cm$^{-3}$) & $N_{e}$
(cm$^{-3}$) & $v_{\textrm{\sc t}}$ (km s$^{-1}$) & $h$ (km) \\
\hline
 -5.19127 &  1584853.9 & 5.145(+00) & 3.474(+08) & 4.168(+08) &  24.997 & 1.07268(+04) \\  
 -5.17478 &  1413985.0 & 7.585(+00) & 4.037(+08) & 4.844(+08) &  24.997 & 7.92048(+03) \\ 
 -5.16217 &  1260642.6 & 1.112(+01) & 4.652(+08) & 5.582(+08) &  24.997 & 6.00415(+03) \\ 
 -5.15270 &  1124094.6 & 1.619(+01) & 5.321(+08) & 6.384(+08) &  24.997 & 4.71803(+03) \\ 
 -5.14560 &  1002513.0 & 2.347(+01) & 6.049(+08) & 7.258(+08) &  24.997 & 3.85491(+03) \\ 
 -5.14027 &   894411.1 & 3.388(+01) & 6.845(+08) & 8.214(+08) &  24.997 & 3.27578(+03) \\ 
 -5.13627 &   797921.2 & 4.885(+01) & 7.721(+08) & 9.263(+08) &  24.997 & 2.88712(+03) \\ 
 -5.13327 &   712247.2 & 7.026(+01) & 8.679(+08) & 1.041(+09) &  24.997 & 2.62625(+03) \\ 
 -5.13103 &   636407.8 & 1.008(+02) & 9.727(+08) & 1.167(+09) &  24.997 & 2.45099(+03) \\ 
 -5.12935 &   568948.6 & 1.444(+02) & 1.088(+09) & 1.305(+09) &  24.997 & 2.33330(+03) \\ 
 -5.12809 &   508595.4 & 2.073(+02) & 1.215(+09) & 1.457(+09) &  24.997 & 2.25409(+03) \\ 
 -5.12714 &   455771.6 & 2.957(+02) & 1.352(+09) & 1.622(+09) &  24.997 & 2.20084(+03) \\ 
 -5.12644 &   409125.0 & 4.206(+02) & 1.500(+09) & 1.800(+09) &  24.997 & 2.16490(+03) \\ 
 -5.12591 &   367575.5 & 5.980(+02) & 1.661(+09) & 1.993(+09) &  24.997 & 2.14060(+03) \\
 -5.12552 &   332266.8 & 8.359(+02) & 1.828(+09) & 2.193(+09) &  24.997 & 2.12413(+03) \\
 -5.12522 &   300036.4 & 1.176(+03) & 2.011(+09) & 2.413(+09) &  24.997 & 2.11294(+03) \\
 -5.12500 &   273347.0 & 1.612(+03) & 2.194(+09) & 2.632(+09) &  24.997 & 2.10531(+03) \\
 -5.12484 &   251069.5 & 2.155(+03) & 2.373(+09) & 2.847(+09) &  24.997 & 2.10005(+03) \\
 -5.12472 &   229606.0 & 2.935(+03) & 2.575(+09) & 3.089(+09) &  24.997 & 2.09643(+03) \\
 -5.12462 &   212574.0 & 3.843(+03) & 2.761(+09) & 3.313(+09) &  24.994 & 2.09391(+03) \\
 -5.12455 &   196161.2 & 5.111(+03) & 2.971(+09) & 3.565(+09) &  24.882 & 2.09191(+03) \\
 -5.12448 &   180448.0 & 6.928(+03) & 3.216(+09) & 3.859(+09) &  24.348 & 2.09025(+03) \\
 -5.12440 &   164226.6 & 9.827(+03) & 3.518(+09) & 4.221(+09) &  23.738 & 2.08871(+03) \\
 -5.12433 &   146786.9 & 1.504(+04) & 3.913(+09) & 4.696(+09) &  23.017 & 2.08732(+03) \\
 -5.12427 &   132660.1 & 2.230(+04) & 4.308(+09) & 5.169(+09) &  22.366 & 2.08618(+03) \\
 -5.12421 &   119174.1 & 3.423(+04) & 4.769(+09) & 5.722(+09) &  21.679 & 2.08525(+03) \\
 -5.12415 &   107764.9 & 5.180(+04) & 5.247(+09) & 6.296(+09) &  21.034 & 2.08441(+03) \\
 -5.12410 &    97241.1 & 8.011(+04) & 5.785(+09) & 6.941(+09) &  20.378 & 2.08364(+03) \\
 -5.12407 &    92216.0 & 1.010(+05) & 6.086(+09) & 7.302(+09) &  20.018 & 2.08329(+03) \\
 -5.12404 &    87313.9 & 1.287(+05) & 6.411(+09) & 7.692(+09) &  19.667 & 2.08296(+03) \\
 -5.12401 &    82287.5 & 1.682(+05) & 6.781(+09) & 8.136(+09) &  19.307 & 2.08265(+03) \\
 -5.12397 &    77765.1 & 2.186(+05) & 7.160(+09) & 8.591(+09) &  18.908 & 2.08226(+03) \\
 -5.12394 &    73242.7 & 2.904(+05) & 7.584(+09) & 9.100(+09) &  18.510 & 2.08188(+03) \\
 -5.12390 &    67996.6 & 4.167(+05) & 8.141(+09) & 9.768(+09) &  18.047 & 2.08153(+03) \\
 -5.12386 &    63590.1 & 5.824(+05) & 8.674(+09) & 1.041(+10) &  17.658 & 2.08120(+03) \\
 -5.12379 &    60973.2 & 7.230(+05) & 9.035(+09) & 1.084(+10) &  17.368 & 2.08060(+03) \\
 -5.12372 &    58494.9 & 8.985(+05) & 9.406(+09) & 1.129(+10) &  17.093 & 2.08003(+03) \\
 -5.12360 &    54142.5 & 1.362(+06) & 1.018(+10) & 1.216(+10) &  16.610 & 2.07917(+03) \\
 -5.12346 &    51072.7 & 1.884(+06) & 1.072(+10) & 1.287(+10) &  16.234 & 2.07817(+03) \\
 -5.12330 &    48242.2 & 2.612(+06) & 1.133(+10) & 1.360(+10) &  15.868 & 2.07709(+03) \\
 -5.12318 &    46212.0 & 3.362(+06) & 1.182(+10) & 1.418(+10) &  15.605 & 2.07638(+03) \\
 -5.12294 &    43432.6 & 4.893(+06) & 1.256(+10) & 1.507(+10) &  15.196 & 2.07490(+03) \\
 -5.12269 &    40763.2 & 7.276(+06) & 1.337(+10) & 1.604(+10) &  14.802 & 2.07351(+03) \\
 -5.12230 &    38241.0 & 1.101(+07) & 1.424(+10) & 1.709(+10) &  14.382 & 2.07144(+03) \\
 -5.12191 &    35751.7 & 1.733(+07) & 1.522(+10) & 1.827(+10) &  13.966 & 2.06950(+03) \\
 -5.12139 &    33983.7 & 2.479(+07) & 1.602(+10) & 1.923(+10) &  13.629 & 2.06706(+03) \\
 -5.12078 &    32054.2 & 3.810(+07) & 1.702(+10) & 2.043(+10) &  13.201 & 2.06431(+03) \\
 -5.12013 &    30042.2 & 6.235(+07) & 1.819(+10) & 2.184(+10) &  12.742 & 2.06159(+03) \\
 -5.11958 &    28359.9 & 9.789(+07) & 1.929(+10) & 2.315(+10) &  12.358 & 2.05943(+03) \\
 -5.11908 &    26829.4 & 1.529(+08) & 2.041(+10) & 2.445(+10) &  12.008 & 2.05761(+03) \\
 -5.11859 &    25301.5 & 2.480(+08) & 2.171(+10) & 2.576(+10) &  11.660 & 2.05589(+03) \\
 -5.11815 &    23939.0 & 3.962(+08) & 2.314(+10) & 2.677(+10) &  11.348 & 2.05445(+03) \\
 -5.11771 &    23198.1 & 5.173(+08) & 2.407(+10) & 2.734(+10) &  11.111 & 2.05309(+03) \\
 -5.11726 &    22448.8 & 6.803(+08) & 2.502(+10) & 2.800(+10) &  10.872 & 2.05177(+03) \\
 -5.11681 &    21697.9 & 8.973(+08) & 2.593(+10) & 2.880(+10) &  10.631 & 2.05050(+03) \\
 -5.11595 &    20262.9 & 1.527(+09) & 2.762(+10) & 3.059(+10) &  10.173 & 2.04822(+03) \\
 -5.11520 &    18997.3 & 2.422(+09) & 2.912(+10) & 3.230(+10) &   9.768 & 2.04636(+03) \\
 -5.11431 &    17510.5 & 4.057(+09) & 3.088(+10) & 3.440(+10) &   9.292 & 2.04437(+03) \\
 -5.11343 &    16023.6 & 6.535(+09) & 3.263(+10) & 3.656(+10) &   8.817 & 2.04257(+03) \\
 -5.11316 &    15474.6 & 7.641(+09) & 3.314(+10) & 3.723(+10) &   8.749 & 2.04207(+03) \\
 -5.11302 &    14826.2 & 8.934(+09) & 3.372(+10) & 3.799(+10) &   8.744 & 2.04183(+03) \\
\end{tabular}
\end{center}
\end{table}

\clearpage

\begin{center}
\begin{tabular}{ccccccc}
\hline
 log $m$ & $T_{e}$ (K) & $N_{H}$ (cm$^{-3}$) & $N_{p}$ (cm$^{-3}$) & $N_{e}$
(cm$^{-3}$) & $v_{\textrm{\sc t}}$ (km s$^{-1}$) & $h$ (km) \\
\hline
 -5.11289 &    14231.2 & 1.021(+10) & 3.428(+10) & 3.872(+10) &   8.741 & 2.04160(+03) \\
 -5.11272 &    13448.8 & 1.201(+10) & 3.508(+10) & 3.973(+10) &   8.736 & 2.04133(+03) \\
 -5.11269 &    13316.1 & 1.234(+10) & 3.523(+10) & 3.994(+10) &   8.735 & 2.04128(+03) \\
 -5.11260 &    12860.5 & 1.348(+10) & 3.574(+10) & 4.059(+10) &   8.733 & 2.04113(+03) \\
 -5.11250 &    12467.5 & 1.454(+10) & 3.621(+10) & 4.113(+10) &   8.731 & 2.04098(+03) \\
 -5.11240 &    12012.6 & 1.582(+10) & 3.682(+10) & 4.174(+10) &   8.728 & 2.04084(+03) \\
 -5.11230 &    11555.1 & 1.718(+10) & 3.759(+10) & 4.220(+10) &   8.725 & 2.04070(+03) \\
 -5.11220 &    11097.5 & 1.861(+10) & 3.862(+10) & 4.233(+10) &   8.723 & 2.04057(+03) \\
 -5.11210 &    10813.9 & 1.958(+10) & 3.937(+10) & 4.220(+10) &   8.720 & 2.04044(+03) \\
 -5.11191 &    10700.7 & 2.016(+10) & 3.959(+10) & 4.203(+10) &   8.719 & 2.04019(+03) \\
 -5.11171 &    10599.0 & 2.071(+10) & 3.978(+10) & 4.188(+10) &   8.718 & 2.03995(+03) \\
 -5.11132 &    10405.0 & 2.176(+10) & 4.012(+10) & 4.164(+10) &   8.717 & 2.03947(+03) \\
 -5.11053 &    10007.5 & 2.393(+10) & 4.075(+10) & 4.142(+10) &   8.713 & 2.03853(+03) \\
 -5.10975 &     9610.0 & 2.620(+10) & 4.123(+10) & 4.151(+10) &   8.710 & 2.03763(+03) \\
 -5.10881 &     9529.3 & 2.727(+10) & 4.107(+10) & 4.131(+10) &   8.702 & 2.03659(+03) \\
 -5.10788 &     9450.1 & 2.830(+10) & 4.093(+10) & 4.114(+10) &   8.694 & 2.03556(+03) \\
 -5.10600 &     9288.7 & 3.029(+10) & 4.072(+10) & 4.088(+10) &   8.678 & 2.03352(+03) \\
 -5.10413 &     9128.8 & 3.225(+10) & 4.054(+10) & 4.068(+10) &   8.661 & 2.03153(+03) \\
 -5.10126 &     8883.4 & 3.522(+10) & 4.034(+10) & 4.046(+10) &   8.637 & 2.02856(+03) \\
 -5.09864 &     8658.3 & 3.793(+10) & 4.022(+10) & 4.032(+10) &   8.614 & 2.02592(+03) \\
 -5.09502 &     8480.8 & 4.077(+10) & 3.996(+10) & 4.006(+10) &   8.593 & 2.02237(+03) \\
 -5.08913 &     8365.4 & 4.402(+10) & 3.949(+10) & 3.960(+10) &   8.571 & 2.01672(+03) \\
 -5.08323 &     8250.0 & 4.721(+10) & 3.912(+10) & 3.922(+10) &   8.550 & 2.01118(+03) \\
 -5.07604 &     8125.0 & 5.094(+10) & 3.872(+10) & 3.883(+10) &   8.525 & 2.00456(+03) \\
 -5.06885 &     8000.0 & 5.468(+10) & 3.839(+10) & 3.850(+10) &   8.500 & 1.99807(+03) \\
 -5.05586 &     7870.0 & 6.057(+10) & 3.787(+10) & 3.799(+10) &   8.460 & 1.98663(+03) \\
 -5.04287 &     7740.0 & 6.649(+10) & 3.749(+10) & 3.762(+10) &   8.420 & 1.97548(+03) \\
 -5.02682 &     7635.0 & 7.371(+10) & 3.729(+10) & 3.742(+10) &   8.320 & 1.96206(+03) \\
 -5.01077 &     7530.0 & 8.108(+10) & 3.726(+10) & 3.739(+10) &   8.220 & 1.94901(+03) \\
 -4.98549 &     7420.3 & 9.204(+10) & 3.745(+10) & 3.760(+10) &   8.099 & 1.92906(+03) \\
 -4.96121 &     7322.3 & 1.031(+11) & 3.787(+10) & 3.804(+10) &   7.976 & 1.91048(+03) \\
 -4.93694 &     7247.5 & 1.147(+11) & 3.850(+10) & 3.867(+10) &   7.834 & 1.89242(+03) \\
 -4.88839 &     7116.9 & 1.399(+11) & 4.006(+10) & 4.026(+10) &   7.577 & 1.85771(+03) \\
 -4.82271 &     7005.8 & 1.773(+11) & 4.242(+10) & 4.265(+10) &   7.323 & 1.81306(+03) \\
 -4.76027 &     6900.2 & 2.207(+11) & 4.477(+10) & 4.503(+10) &   7.081 & 1.77253(+03) \\
 -4.66481 &     6786.4 & 3.030(+11) & 4.839(+10) & 4.871(+10) &   6.733 & 1.71400(+03) \\
 -4.55600 &     6698.4 & 4.275(+11) & 5.215(+10) & 5.260(+10) &   6.357 & 1.65133(+03) \\
 -4.43506 &     6600.7 & 6.211(+11) & 5.566(+10) & 5.618(+10) &   5.938 & 1.58601(+03) \\
 -4.30139 &     6535.7 & 9.185(+11) & 5.898(+10) & 5.970(+10) &   5.523 & 1.51842(+03) \\
 -4.15567 &     6475.9 & 1.385(+12) & 6.294(+10) & 6.392(+10) &   5.096 & 1.44897(+03) \\
 -4.07753 &     6443.9 & 1.718(+12) & 6.524(+10) & 6.640(+10) &   4.877 & 1.41292(+03) \\
 -3.99939 &     6411.8 & 2.125(+12) & 6.812(+10) & 6.950(+10) &   4.659 & 1.37785(+03) \\
 -3.91534 &     6380.7 & 2.661(+12) & 7.222(+10) & 7.387(+10) &   4.432 & 1.34114(+03) \\
 -3.83129 &     6355.5 & 3.320(+12) & 7.739(+10) & 7.937(+10) &   4.219 & 1.30536(+03) \\
 -3.65298 &     6256.8 & 5.324(+12) & 8.481(+10) & 8.769(+10) &   3.777 & 1.23347(+03) \\
 -3.47354 &     6127.5 & 8.547(+12) & 8.741(+10) & 9.159(+10) &   3.356 & 1.16498(+03) \\
 -3.30000 &     6135.4 & 1.313(+13) & 1.125(+11) & 1.187(+11) &   2.960 & 1.10117(+03) \\
 -3.13428 &     6112.0 & 1.982(+13) & 1.356(+11) & 1.445(+11) &   2.592 & 1.04186(+03) \\
 -2.97768 &     6044.7 & 2.936(+13) & 1.489(+11) & 1.612(+11) &   2.262 & 9.87388(+02) \\
 -2.83040 &     5955.0 & 4.258(+13) & 1.539(+11) & 1.702(+11) &   1.961 & 9.37643(+02) \\
 -2.68992 &     5851.9 & 6.071(+13) & 1.520(+11) & 1.729(+11) &   1.689 & 8.91579(+02) \\
 -2.55407 &     5736.4 & 8.529(+13) & 1.433(+11) & 1.690(+11) &   1.538 & 8.48232(+02) \\
 -2.42406 &     5589.2 & 1.187(+14) & 1.230(+11) & 1.527(+11) &   1.416 & 8.07900(+02) \\
 -2.29839 &     5444.7 & 1.635(+14) & 1.027(+11) & 1.365(+11) &   1.299 & 7.70060(+02) \\
 -2.17917 &     5291.1 & 2.222(+14) & 8.018(+10) & 1.175(+11) &   1.193 & 7.35239(+02) \\
 -2.07437 &     5132.7 & 2.923(+14) & 5.753(+10) & 9.920(+10) &   1.108 & 7.05562(+02) \\
 -2.02936 &     5061.5 & 3.291(+14) & 4.829(+10) & 9.219(+10) &   1.073 & 6.93045(+02) \\
\end{tabular}
\end{center}

\clearpage

\begin{center}
\begin{tabular}{ccccccc}
\hline
 log $m$ & $T_{e}$ (K) & $N_{H}$ (cm$^{-3}$) & $N_{p}$ (cm$^{-3}$) & $N_{e}$
(cm$^{-3}$) & $v_{\textrm{\sc t}}$ (km s$^{-1}$) & $h$ (km) \\
\hline
 -1.98435 &     4987.5 & 3.708(+14) & 3.945(+10) & 8.598(+10) &   1.041 & 6.80717(+02) \\
 -1.90597 &     4858.6 & 4.566(+14) & 2.616(+10) & 7.875(+10) &   0.985 & 6.59769(+02) \\
 -1.83566 &     4746.6 & 5.502(+14) & 1.718(+10) & 7.715(+10) &   0.937 & 6.41445(+02) \\
 -1.77067 &     4646.6 & 6.534(+14) & 1.120(+10) & 7.977(+10) &   0.894 & 6.24892(+02) \\
 -1.70766 &     4549.5 & 7.723(+14) & 7.059(+09) & 8.551(+10) &   0.851 & 6.09192(+02) \\
 -1.64196 &     4469.2 & 9.155(+14) & 4.747(+09) & 9.481(+10) &   0.809 & 5.93155(+02) \\
 -1.56959 &     4402.8 & 1.099(+15) & 3.421(+09) & 1.076(+11) &   0.763 & 5.75802(+02) \\
 -1.48889 &     4328.7 & 1.348(+15) & 2.345(+09) & 1.231(+11) &   0.713 & 5.56793(+02) \\
 -1.39929 &     4297.7 & 1.670(+15) & 2.089(+09) & 1.450(+11) &   0.669 & 5.35970(+02) \\
 -1.30098 &     4280.3 & 2.105(+15) & 2.023(+09) & 1.741(+11) &   0.626 & 5.13289(+02) \\
 -1.18674 &     4277.6 & 2.742(+15) & 2.147(+09) & 2.167(+11) &   0.582 & 4.87053(+02) \\
 -1.05256 &     4295.8 & 3.722(+15) & 2.574(+09) & 2.837(+11) &   0.540 & 4.56269(+02) \\
 -0.90681 &     4354.9 & 5.137(+15) & 3.845(+09) & 3.911(+11) &   0.527 & 4.22575(+02) \\
 -0.76297 &     4423.9 & 7.043(+15) & 5.907(+09) & 5.405(+11) &   0.524 & 3.88815(+02) \\
 -0.62093 &     4492.0 & 9.620(+15) & 8.866(+09) & 7.427(+11) &   0.520 & 3.54964(+02) \\
 -0.48037 &     4586.4 & 1.302(+16) & 1.511(+10) & 1.034(+12) &   0.555 & 3.20827(+02) \\
 -0.34126 &     4683.4 & 1.755(+16) & 2.612(+10) & 1.434(+12) &   0.593 & 2.86306(+02) \\
 -0.20325 &     4782.3 & 2.359(+16) & 4.641(+10) & 1.981(+12) &   0.637 & 2.51307(+02) \\
 -0.06709 &     4930.9 & 3.122(+16) & 1.088(+11) & 2.798(+12) &   0.774 & 2.15775(+02) \\
  0.03467 &     5042.0 & 3.852(+16) & 2.104(+11) & 3.626(+12) &   0.877 & 1.88360(+02) \\
  0.11629 &     5131.1 & 4.559(+16) & 3.582(+11) & 4.480(+12) &   0.959 & 1.65848(+02) \\
  0.18791 &     5225.7 & 5.268(+16) & 6.114(+11) & 5.477(+12) &   1.037 & 1.45685(+02) \\
  0.25064 &     5327.2 & 5.959(+16) & 1.043(+12) & 6.671(+12) &   1.111 & 1.27645(+02) \\
  0.30637 &     5417.4 & 6.650(+16) & 1.647(+12) & 8.034(+12) &   1.176 & 1.11293(+02) \\
  0.35814 &     5527.4 & 7.330(+16) & 2.721(+12) & 9.911(+12) &   1.240 & 9.57905(+01) \\
  0.41162 &     5658.4 & 8.083(+16) & 4.692(+12) & 1.281(+13) &   1.307 & 7.93916(+01) \\
  0.46173 &     5781.2 & 8.863(+16) & 7.524(+12) & 1.659(+13) &   1.370 & 6.36483(+01) \\
  0.50584 &     5919.9 & 9.566(+16) & 1.208(+13) & 2.209(+13) &   1.428 & 4.94471(+01) \\
  0.54316 &     6068.0 & 1.016(+17) & 1.897(+13) & 2.987(+13) &   1.479 & 3.71186(+01) \\
  0.57682 &     6201.6 & 1.073(+17) & 2.766(+13) & 3.944(+13) &   1.525 & 2.57236(+01) \\
  0.61048 &     6335.2 & 1.133(+17) & 3.936(+13) & 5.205(+13) &   1.571 & 1.40667(+01) \\
  0.64141 &     6490.1 & 1.186(+17) & 5.715(+13) & 7.088(+13) &   1.614 & 3.09627(+00) \\
  0.65687 &     6603.4 & 1.208(+17) & 7.334(+13) & 8.765(+13) &   1.637 & -2.50818(+00) \\
  0.67234 &     6716.7 & 1.229(+17) & 9.314(+13) & 1.081(+14) &   1.661 & -8.21229(+00) \\
  0.69582 &     6888.8 & 1.264(+17) & 1.315(+14) & 1.475(+14) &   1.696 & -1.70655(+01) \\
  0.70757 &     7011.2 & 1.275(+17) & 1.651(+14) & 1.817(+14) &   1.709 & -2.15903(+01) \\
  0.71931 &     7145.4 & 1.285(+17) & 2.095(+14) & 2.269(+14) &   1.720 & -2.61991(+01) \\
  0.73755 &     7353.9 & 1.301(+17) & 2.977(+14) & 3.164(+14) &   1.738 & -3.35360(+01) \\
  0.75580 &     7562.4 & 1.318(+17) & 4.146(+14) & 4.346(+14) &   1.756 & -4.10870(+01) \\
  0.77133 &     7754.9 & 1.330(+17) & 5.52E(+14) & 5.744(+14) &   1.768 & -4.76907(+01) \\
  0.78686 &     7953.0 & 1.341(+17) & 7.323(+14) & 7.554(+14) &   1.779 & -5.44673(+01) \\
  0.80107 &     8134.1 & 1.351(+17) & 9.357(+14) & 9.603(+14) &   1.790 & -6.08198(+01) \\
  0.81517 &     8313.9 & 1.362(+17) & 1.181(+15) & 1.207(+15) &   1.800 & -6.72742(+01) \\
\hline
\end{tabular}
\end{center}

\clearpage

\begin{table}
\caption{Network model X}\label{tabA8}
\begin{center}
\begin{tabular}{ccccccc}
\hline
 log $m$ & $T_{e}$ (K) & $N_{H}$ (cm$^{-3}$) & $N_{p}$ (cm$^{-3}$) & $N_{e}$
(cm$^{-3}$) & $v_{\textrm{\sc t}}$ (km s$^{-1}$) & $h$ (km) \\
\hline
 -5.19116 &  1584822.0 & 5.146(+00) & 3.474(+08) & 4.169(+08) &   24.997 & 1.07156(+04) \\ 
 -5.17478 &  1415053.0 & 7.568(+00) & 4.034(+08) & 4.840(+08) &   24.997 & 7.92667(+03) \\ 
 -5.16217 &  1261894.6 & 1.108(+01) & 4.648(+08) & 5.577(+08) &   24.997 & 6.00867(+03) \\ 
 -5.15270 &  1125551.6 & 1.613(+01) & 5.314(+08) & 6.376(+08) &   24.997 & 4.72109(+03) \\
 -5.14560 &  1004223.0 & 2.334(+01) & 6.039(+08) & 7.246(+08) &   24.997 & 3.85669(+03) \\ 
 -5.14027 &   896382.5 & 3.365(+01) & 6.831(+08) & 8.196(+08) &   24.997 & 3.27644(+03) \\ 
 -5.13627 &   800258.4 & 4.840(+01) & 7.699(+08) & 9.237(+08) &   24.997 & 2.88681(+03) \\ 
 -5.13327 &   715009.6 & 6.941(+01) & 8.647(+08) & 1.038(+09) &   24.997 & 2.62507(+03) \\ 
 -5.13103 &   639581.0 & 9.920(+01) & 9.681(+08) & 1.162(+09) &   24.997 & 2.44905(+03) \\ 
 -5.12935 &   572675.9 & 1.414(+02) & 1.081(+09) & 1.297(+09) &   24.997 & 2.33070(+03) \\ 
 -5.12809 &   513098.5 & 2.015(+02) & 1.204(+09) & 1.445(+09) &   24.997 & 2.25091(+03) \\ 
 -5.12714 &   461102.6 & 2.848(+02) & 1.337(+09) & 1.604(+09) &   24.997 & 2.19713(+03) \\ 
 -5.12644 &   415441.6 & 4.001(+02) & 1.478(+09) & 1.774(+09) &   24.997 & 2.16072(+03) \\ 
 -5.12591 &   375127.7 & 5.594(+02) & 1.630(+09) & 1.955(+09) &   24.997 & 2.13602(+03) \\ 
 -5.12552 &   341028.9 & 7.667(+02) & 1.784(+09) & 2.140(+09) &   24.997 & 2.11918(+03) \\ 
 -5.12522 &   310784.0 & 1.045(+03) & 1.947(+09) & 2.336(+09) &   24.997 & 2.10766(+03) \\ 
 -5.12500 &   285294.0 & 1.395(+03) & 2.109(+09) & 2.530(+09) &   24.997 & 2.09975(+03) \\ 
 -5.12484 &   263687.9 & 1.823(+03) & 2.268(+09) & 2.722(+09) &   24.997 & 2.09426(+03) \\ 
 -5.12472 &   244839.8 & 2.350(+03) & 2.428(+09) & 2.914(+09) &   24.997 & 2.09044(+03) \\ 
 -5.12462 &   229435.0 & 2.943(+03) & 2.577(+09) & 3.092(+09) &   24.997 & 2.08776(+03) \\ 
 -5.12455 &   214725.0 & 3.710(+03) & 2.737(+09) & 3.284(+09) &   24.997 & 2.08561(+03) \\ 
 -5.12448 &   201497.3 & 4.643(+03) & 2.898(+09) & 3.477(+09) &   24.997 & 2.08378(+03) \\ 
 -5.12440 &   185830.1 & 6.224(+03) & 3.129(+09) & 3.754(+09) &   24.528 & 2.08206(+03) \\ 
 -5.12433 &   168687.5 & 8.893(+03) & 3.430(+09) & 4.116(+09) &   23.908 & 2.08048(+03) \\ 
 -5.12427 &   153297.0 & 1.275(+04) & 3.756(+09) & 4.507(+09) &   23.294 & 2.07918(+03) \\ 
 -5.12421 &   137634.8 & 1.931(+04) & 4.161(+09) & 4.992(+09) &   22.607 & 2.07812(+03) \\ 
 -5.12415 &   123210.5 & 2.993(+04) & 4.621(+09) & 5.545(+09) &   21.897 & 2.07716(+03) \\ 
 -5.12410 &   109670.1 & 4.815(+04) & 5.161(+09) & 6.192(+09) &   21.151 & 2.07628(+03) \\ 
 -5.12407 &   102924.4 & 6.285(+04) & 5.482(+09) & 6.577(+09) &   20.742 & 2.07590(+03) \\ 
 -5.12404 &    96492.7 & 8.285(+04) & 5.828(+09) & 6.993(+09) &   20.328 & 2.07554(+03) \\ 
 -5.12401 &    89901.9 & 1.129(+05) & 6.233(+09) & 7.479(+09) &   19.880 & 2.07520(+03) \\ 
 -5.12397 &    83855.4 & 1.544(+05) & 6.661(+09) & 7.992(+09) &   19.426 & 2.07477(+03) \\ 
 -5.12394 &    78651.6 & 2.073(+05) & 7.080(+09) & 8.495(+09) &   19.019 & 2.07437(+03) \\ 
 -5.12390 &    73817.0 & 2.796(+05) & 7.522(+09) & 9.026(+09) &   18.608 & 2.07399(+03) \\ 
 -5.12386 &    70257.6 & 3.548(+05) & 7.885(+09) & 9.461(+09) &   18.299 & 2.07363(+03) \\ 
 -5.12379 &    65101.5 & 5.175(+05) & 8.488(+09) & 1.018(+10) &   17.803 & 2.07298(+03) \\ 
 -5.12372 &    61381.1 & 6.986(+05) & 8.977(+09) & 1.077(+10) &   17.426 & 2.07238(+03) \\ 
 -5.12362 &    56778.8 & 1.053(+06) & 9.676(+09) & 1.161(+10) &   16.927 & 2.07164(+03) \\ 
 -5.12354 &    53044.4 & 1.525(+06) & 1.033(+10) & 1.240(+10) &   16.483 & 2.07104(+03) \\ 
 -5.12346 &    49259.7 & 2.315(+06) & 1.110(+10) & 1.332(+10) &   16.018 & 2.07048(+03) \\ 
 -5.12338 &    45588.0 & 3.648(+06) & 1.197(+10) & 1.437(+10) &   15.498 & 2.06997(+03) \\ 
 -5.12330 &    41992.1 & 6.031(+06) & 1.297(+10) & 1.556(+10) &   14.988 & 2.06949(+03) \\ 
 -5.12321 &    38304.7 & 1.106(+07) & 1.421(+10) & 1.705(+10) &   14.389 & 2.06904(+03) \\ 
 -5.12312 &    34644.4 & 2.158(+07) & 1.566(+10) & 1.880(+10) &   13.743 & 2.06860(+03) \\ 
 -5.12306 &    32255.3 & 3.630(+07) & 1.680(+10) & 2.017(+10) &   13.302 & 2.06832(+03) \\ 
 -5.12300 &    29930.5 & 6.410(+07) & 1.810(+10) & 2.172(+10) &   12.805 & 2.06806(+03) \\ 
 -5.12294 &    27636.5 & 1.207(+08) & 1.958(+10) & 2.348(+10) &   12.310 & 2.06783(+03) \\ 
 -5.12288 &    25957.6 & 2.029(+08) & 2.087(+10) & 2.491(+10) &   11.891 & 2.06761(+03) \\ 
 -5.12282 &    24080.4 & 3.888(+08) & 2.269(+10) & 2.635(+10) &   11.423 & 2.06740(+03) \\ 
 -5.12275 &    22773.8 & 6.398(+08) & 2.422(+10) & 2.727(+10) &   11.055 & 2.06722(+03) \\ 
 -5.12269 &    21590.9 & 1.025(+09) & 2.553(+10) & 2.835(+10) &   10.722 & 2.06704(+03) \\ 
 -5.12260 &    19954.5 & 2.009(+09) & 2.720(+10) & 3.016(+10) &   10.227 & 2.06678(+03) \\ 
 -5.12250 &    19150.9 & 2.759(+09) & 2.801(+10) & 3.111(+10) &   9.954 & 2.06654(+03) \\ 
 -5.12230 &    17636.9 & 4.765(+09) & 2.947(+10) & 3.292(+10) &   9.432 & 2.06610(+03) \\ 
 -5.12211 &    17553.1 & 4.613(+09) & 2.973(+10) & 3.319(+10) &   9.413 & 2.06569(+03) \\ 
 -5.12191 &    17482.5 & 4.507(+09) & 2.994(+10) & 3.340(+10) &   9.399 & 2.06527(+03) \\ 
 -5.12165 &    17388.2 & 4.428(+09) & 3.018(+10) & 3.366(+10) &   9.380 & 2.06472(+03) \\ 
 -5.12139 &    17294.2 & 4.391(+09) & 3.039(+10) & 3.389(+10) &   9.361 & 2.06417(+03) \\ 
 -5.12078 &    17072.8 & 4.398(+09) & 3.085(+10) & 3.439(+10) &   9.316 & 2.06287(+03) \\ 
 -5.12013 &    16837.9 & 4.494(+09) & 3.129(+10) & 3.488(+10) &   9.269 & 2.06151(+03) \\ 
 -5.11958 &    16641.7 & 4.616(+09) & 3.164(+10) & 3.528(+10) &   9.229 & 2.06039(+03) \\ 
\end{tabular}
\end{center}
\end{table}

\clearpage

\begin{center}
\begin{tabular}{ccccccc}
\hline
 log $m$ & $T_{e}$ (K) & $N_{H}$ (cm$^{-3}$) & $N_{p}$ (cm$^{-3}$) & $N_{e}$
(cm$^{-3}$) & $v_{\textrm{\sc t}}$ (km s$^{-1}$) & $h$ (km) \\
\hline
 -5.11908 &    16463.1 & 4.756(+09) & 3.195(+10) & 3.563(+10) &   9.193 & 2.05939(+03) \\ 
 -5.11815 &    16125.6 & 5.086(+09) & 3.251(+10) & 3.629(+10) &   9.124 & 2.05752(+03) \\ 
 -5.11726 &    15805.2 & 5.471(+09) & 3.304(+10) & 3.690(+10) &   9.059 & 2.05579(+03) \\ 
 -5.11595 &    15335.3 & 6.176(+09) & 3.377(+10) & 3.777(+10) &   8.964 & 2.05331(+03) \\ 
 -5.11431 &    14743.6 & 7.366(+09) & 3.460(+10) & 3.879(+10) &   8.844 & 2.05032(+03) \\ 
 -5.11316 &    14329.0 & 8.504(+09) & 3.506(+10) & 3.941(+10) &   8.761 & 2.04830(+03) \\ 
 -5.11302 &    14278.7 & 8.671(+09) & 3.511(+10) & 3.947(+10) &   8.750 & 2.04807(+03) \\ 
 -5.11289 &    14231.2 & 8.839(+09) & 3.514(+10) & 3.952(+10) &   8.741 & 2.04784(+03) \\ 
 -5.11272 &    13426.1 & 1.040(+10) & 3.615(+10) & 4.078(+10) &   8.736 & 2.04756(+03) \\ 
 -5.11269 &    13299.7 & 1.066(+10) & 3.632(+10) & 4.098(+10) &   8.735 & 2.04751(+03) \\ 
 -5.11260 &    12840.8 & 1.164(+10) & 3.695(+10) & 4.173(+10) &   8.733 & 2.04736(+03) \\ 
 -5.11250 &    12493.8 & 1.245(+10) & 3.745(+10) & 4.230(+10) &   8.731 & 2.04721(+03) \\ 
 -5.11240 &    12037.0 & 1.355(+10) & 3.819(+10) & 4.302(+10) &   8.728 & 2.04706(+03) \\ 
 -5.11230 &    11571.5 & 1.475(+10) & 3.908(+10) & 4.361(+10) &   8.725 & 2.04693(+03) \\ 
 -5.11220 &    11105.7 & 1.601(+10) & 4.025(+10) & 4.387(+10) &   8.723 & 2.04679(+03) \\ 
 -5.11210 &    10813.9 & 1.687(+10) & 4.111(+10) & 4.384(+10) &   8.720 & 2.04666(+03) \\ 
 -5.11191 &    10698.9 & 1.735(+10) & 4.141(+10) & 4.375(+10) &   8.719 & 2.04641(+03) \\ 
 -5.11171 &    10596.2 & 1.779(+10) & 4.167(+10) & 4.368(+10) &   8.718 & 2.04616(+03) \\ 
 -5.11132 &    10452.9 & 1.849(+10) & 4.199(+10) & 4.357(+10) &   8.716 & 2.04567(+03) \\ 
 -5.11053 &    10280.9 & 1.954(+10) & 4.228(+10) & 4.341(+10) &   8.710 & 2.04470(+03) \\ 
 -5.10975 &    10109.0 & 2.060(+10) & 4.254(+10) & 4.333(+10) &   8.705 & 2.04375(+03) \\ 
 -5.10881 &     9903.3 & 2.189(+10) & 4.282(+10) & 4.332(+10) &   8.698 & 2.04264(+03) \\ 
 -5.10788 &     9699.1 & 2.321(+10) & 4.306(+10) & 4.338(+10) &   8.691 & 2.04156(+03) \\ 
 -5.10600 &     9288.2 & 2.598(+10) & 4.349(+10) & 4.365(+10) &   8.678 & 2.03946(+03) \\ 
 -5.10413 &     9128.8 & 2.769(+10) & 4.348(+10) & 4.361(+10) &   8.661 & 2.03742(+03) \\ 
 -5.10126 &     8883.6 & 3.032(+10) & 4.350(+10) & 4.361(+10) &   8.637 & 2.03438(+03) \\ 
 -5.09864 &     8658.1 & 3.276(+10) & 4.356(+10) & 4.367(+10) &   8.614 & 2.03168(+03) \\ 
 -5.09502 &     8480.8 & 3.533(+10) & 4.349(+10) & 4.351(+10) &   8.593 & 2.02804(+03) \\ 
 -5.08913 &     8365.4 & 3.831(+10) & 4.320(+10) & 4.331(+10) &   8.571 & 2.02225(+03) \\ 
 -5.08323 &     8250.0 & 4.126(+10) & 4.298(+10) & 4.309(+10) &   8.550 & 2.01657(+03) \\ 
 -5.07604 &     8125.0 & 4.476(+10) & 4.274(+10) & 4.285(+10) &   8.525 & 2.00979(+03) \\ 
 -5.07245 &     8062.5 & 4.654(+10) & 4.263(+10) & 4.274(+10) &   8.512 & 2.00645(+03) \\ 
 -5.06885 &     8000.0 & 4.832(+10) & 4.254(+10) & 4.265(+10) &   8.500 & 2.00314(+03) \\ 
 -5.05586 &     7870.0 & 5.399(+10) & 4.217(+10) & 4.228(+10) &   8.460 & 1.99143(+03) \\ 
 -5.04287 &     7740.0 & 5.978(+10) & 4.188(+10) & 4.200(+10) &   8.420 & 1.98002(+03) \\ 
 -5.02682 &     7635.0 & 6.690(+10) & 4.173(+10) & 4.186(+10) &   8.320 & 1.96630(+03) \\ 
 -5.01077 &     7530.0 & 7.422(+10) & 4.173(+10) & 4.186(+10) &   8.220 & 1.95298(+03) \\ 
 -4.98543 &     7420.3 & 8.517(+10) & 4.192(+10) & 4.206(+10) &   8.099 & 1.93263(+03) \\ 
 -4.96121 &     7322.3 & 9.622(+10) & 4.233(+10) & 4.249(+10) &   7.976 & 1.91371(+03) \\ 
 -4.93694 &     7247.4 & 1.079(+11) & 4.295(+10) & 4.311(+10) &   7.834 & 1.89535(+03) \\ 
 -4.88839 &     7116.9 & 1.330(+11) & 4.452(+10) & 4.471(+10) &   7.577 & 1.86012(+03) \\ 
 -4.82271 &     7005.9 & 1.704(+11) & 4.687(+10) & 4.709(+10) &   7.323 & 1.81490(+03) \\ 
 -4.76027 &     6900.2 & 2.138(+11) & 4.918(+10) & 4.943(+10) &   7.081 & 1.77394(+03) \\ 
 -4.66481 &     6786.4 & 2.962(+11) & 5.261(+10) & 5.293(+10) &   6.733 & 1.71493(+03) \\ 
 -4.55600 &     6698.4 & 4.214(+11) & 5.595(+10) & 5.636(+10) &   6.357 & 1.65188(+03) \\ 
 -4.43506 &     6600.7 & 6.164(+11) & 5.857(+10) & 5.909(+10) &   5.938 & 1.58639(+03) \\ 
 -4.30139 &     6535.7 & 9.154(+11) & 6.080(+10) & 6.152(+10) &   5.523 & 1.51861(+03) \\ 
 -4.15567 &     6475.9 & 1.383(+12) & 6.411(+10) & 6.509(+10) &   5.096 & 1.44911(+03) \\ 
 -4.07753 &     6443.9 & 1.716(+12) & 6.619(+10) & 6.735(+10) &   4.877 & 1.41303(+03) \\ 
 -3.99939 &     6411.8 & 2.123(+12) & 6.888(+10) & 7.025(+10) &   4.659 & 1.37795(+03) \\ 
 -3.91534 &     6380.7 & 2.660(+12) & 7.303(+10) & 7.468(+10) &   4.432 & 1.34124(+03) \\ 
 -3.83129 &     6355.5 & 3.318(+12) & 7.839(+10) & 8.037(+10) &   4.219 & 1.30545(+03) \\ 
 -3.65298 &     6256.8 & 5.318(+12) & 8.810(+10) & 9.096(+10) &   3.777 & 1.23355(+03) \\ 
 -3.47354 &     6127.5 & 8.539(+12) & 9.151(+10) & 9.567(+10) &   3.356 & 1.16501(+03) \\ 
 -3.30000 &     6135.4 & 1.313(+13) & 1.141(+11) & 1.203(+11) &   2.960 & 1.10119(+03) \\ 
 -3.13428 &     6112.0 & 1.982(+13) & 1.376(+11) & 1.465(+11) &   2.592 & 1.04187(+03) \\ 
 -2.97768 &     6044.7 & 2.936(+13) & 1.514(+11) & 1.637(+11) &   2.262 & 9.87399(+02) \\ 
\end{tabular}
\end{center}

\clearpage

\begin{center}
\begin{tabular}{ccccccc}
\hline
 log $m$ & $T_{e}$ (K) & $N_{H}$ (cm$^{-3}$) & $N_{p}$ (cm$^{-3}$) & $N_{e}$
(cm$^{-3}$) & $v_{\textrm{\sc t}}$ (km s$^{-1}$) & $h$ (km) \\
\hline
 -2.83040 &     5955.0 & 4.257(+13) & 1.566(+11) & 1.729(+11) &   1.961 & 9.37651(+02) \\ 
 -2.68992 &     5851.9 & 6.070(+13) & 1.550(+11) & 1.759(+11) &   1.689 & 8.91585(+02) \\ 
 -2.55407 &     5736.4 & 8.528(+13) & 1.467(+11) & 1.722(+11) &   1.538 & 8.48236(+02) \\ 
 -2.42406 &     5589.2 & 1.187(+14) & 1.274(+11) & 1.569(+11) &   1.416 & 8.07902(+02) \\ 
 -2.29839 &     5444.7 & 1.635(+14) & 1.076(+11) & 1.410(+11) &   1.299 & 7.70062(+02) \\ 
 -2.17917 &     5291.1 & 2.222(+14) & 8.586(+10) & 1.230(+11) &   1.193 & 7.35240(+02) \\ 
 -2.07437 &     5132.7 & 2.923(+14) & 6.353(+10) & 1.047(+11) &   1.108 & 7.05562(+02) \\ 
 -2.02936 &     5061.5 & 3.291(+14) & 5.088(+10) & 9.455(+10) &   1.073 & 6.93045(+02) \\ 
 -1.98435 &     4987.5 & 3.708(+14) & 4.206(+10) & 8.838(+10) &   1.041 & 6.80717(+02) \\ 
 -1.90597 &     4858.6 & 4.566(+14) & 3.061(+10) & 8.291(+10) &   0.985 & 6.59768(+02) \\ 
 -1.83566 &     4746.6 & 5.502(+14) & 2.052(+10) & 8.032(+10) &   0.937 & 6.41444(+02) \\ 
 -1.77067 &     4646.6 & 6.534(+14) & 1.359(+10) & 8.206(+10) &   0.894 & 6.24891(+02) \\ 
 -1.70766 &     4549.5 & 7.723(+14) & 8.764(+09) & 8.713(+10) &   0.851 & 6.09191(+02) \\ 
 -1.64196 &     4469.2 & 9.155(+14) & 5.754(+09) & 9.576(+10) &   0.809 & 5.93154(+02) \\ 
 -1.56959 &     4402.8 & 1.099(+15) & 4.027(+09) & 1.081(+11) &   0.763 & 5.75801(+02) \\ 
 -1.48889 &     4328.7 & 1.348(+15) & 2.794(+09) & 1.235(+11) &   0.713 & 5.56793(+02) \\ 
 -1.39929 &     4297.7 & 1.670(+15) & 2.267(+09) & 1.451(+11) &   0.669 & 5.35970(+02) \\ 
 -1.30098 &     4280.3 & 2.105(+15) & 2.136(+09) & 1.742(+11) &   0.626 & 5.13289(+02) \\ 
 -1.18674 &     4277.6 & 2.742(+15) & 2.223(+09) & 2.168(+11) &   0.584 & 4.87053(+02) \\ 
 -1.05256 &     4295.8 & 3.722(+15) & 2.624(+09) & 2.837(+11) &   0.540 & 4.56269(+02) \\ 
 -0.90681 &     4354.9 & 5.137(+15) & 3.876(+09) & 3.911(+11) &   0.527 & 4.22570(+02) \\ 
 -0.76297 &     4423.9 & 7.043(+15) & 5.966(+09) & 5.406(+11) &   0.524 & 3.88814(+02) \\ 
 -0.62093 &     4492.0 & 9.620(+15) & 8.962(+09) & 7.427(+11) &   0.520 & 3.54964(+02) \\ 
 -0.48037 &     4586.4 & 1.302(+16) & 1.527(+10) & 1.034(+12) &   0.555 & 3.20827(+02) \\ 
 -0.34126 &     4683.4 & 1.755(+16) & 2.637(+10) & 1.434(+12) &   0.593 & 2.86306(+02) \\ 
 -0.20325 &     4782.3 & 2.359(+16) & 4.675(+10) & 1.982(+12) &   0.637 & 2.51306(+02) \\ 
 -0.06709 &     4930.9 & 3.122(+16) & 1.092(+11) & 2.798(+12) &   0.774 & 2.15774(+02) \\ 
  0.03467 &     5042.0 & 3.852(+16) & 2.108(+11) & 3.626(+12) &   0.877 & 1.88359(+02) \\ 
  0.11629 &     5131.1 & 4.559(+16) & 3.585(+11) & 4.480(+12) &   0.959 & 1.65848(+02) \\ 
  0.18791 &     5225.7 & 5.268(+16) & 6.116(+11) & 5.477(+12) &   1.037 & 1.45685(+02) \\ 
  0.25064 &     5327.2 & 5.959(+16) & 1.043(+12) & 6.671(+12) &   1.111 & 1.27645(+02) \\ 
  0.30637 &     5417.4 & 6.650(+16) & 1.647(+12) & 8.035(+12) &   1.176 & 1.11292(+02) \\ 
  0.35814 &     5527.4 & 7.330(+16) & 2.721(+12) & 9.911(+12) &   1.240 & 9.57900(+01) \\ 
  0.41162 &     5658.4 & 8.083(+16) & 4.692(+12) & 1.281(+13) &   1.307 & 7.93911(+01) \\
  0.46173 &     5781.2 & 8.863(+16) & 7.524(+12) & 1.659(+13) &   1.370 & 6.36477(+01) \\
  0.50584 &     5919.9 & 9.566(+16) & 1.208(+13) & 2.209(+13) &   1.428 & 4.94465(+01) \\
  0.54316 &     6068.0 & 1.016(+17) & 1.897(+13) & 2.987(+13) &   1.479 & 3.71181(+01) \\
  0.57682 &     6201.6 & 1.073(+17) & 2.766(+13) & 3.944(+13) &   1.525 & 2.57231(+01) \\
  0.61048 &     6335.2 & 1.133(+17) & 3.936(+13) & 5.209(+13) &   1.571 & 1.40661(+01) \\
  0.64141 &     6490.1 & 1.186(+17) & 5.715(+13) & 7.088(+13) &   1.614 & 3.09574(+00) \\
  0.65687 &     6603.4 & 1.208(+17) & 7.334(+13) & 8.765(+13) &   1.637 & -2.5087(+00) \\
  0.67234 &     6716.7 & 1.229(+17) & 9.314(+13) & 1.081(+14) &   1.661 & -8.21282(+00) \\
  0.69582 &     6888.8 & 1.264(+17) & 1.315(+14) & 1.475(+14) &   1.696 & -1.70661(+01) \\
  0.70757 &     7011.2 & 1.275(+17) & 1.651(+14) & 1.817(+14) &   1.709 & -2.15908(+01) \\
  0.71931 &     7145.4 & 1.285(+17) & 2.095(+14) & 2.269(+14) &   1.720 & -2.61997(+01) \\
  0.73755 &     7353.9 & 1.301(+17) & 2.977(+14) & 3.164(+14) &   1.738 & -3.35366(+01) \\
  0.75580 &     7562.4 & 1.318(+17) & 4.146(+14) & 4.346(+14) &   1.756 & -4.10876(+01) \\
  0.77133 &     7754.9 & 1.330(+17) & 5.529(+14) & 5.744(+14) &   1.768 & -4.76913(+01) \\
  0.78686 &     7953.0 & 1.341(+17) & 7.323(+14) & 7.553(+14) &   1.779 & -5.44678(+01) \\
  0.80107 &     8134.1 & 1.351(+17) & 9.357(+14) & 9.603(+14) &   1.790 & -6.08203(+01) \\
  0.81517 &     8313.9 & 1.362(+17) & 1.181(+15) & 1.207(+15) &   1.800 & -6.72747(+01) \\
\hline
\end{tabular}
\end{center}

\end{document}